\documentclass[onecolumn,12pt,journal,draftclsnofoot]{IEEEtran}
\usepackage[latin9]{inputenc}
\usepackage{geometry}
\usepackage{float}
\usepackage{amsmath}
\usepackage{amssymb}
\usepackage{graphicx}

\makeatletter

\floatstyle{ruled}
\newfloat{algorithm}{tbp}{loa}
\providecommand{\algorithmname}{Algorithm}
\floatname{algorithm}{\protect\algorithmname}

\usepackage{amsfonts}
\usepackage{mathrsfs}
\usepackage{mathrsfs}\usepackage[noblocks]{authblk}
\usepackage[compress]{cite}
\usepackage{bm}
\usepackage{algorithm}
\usepackage{algorithmic}
\usepackage{epsfig}\usepackage{graphics}\usepackage{subfigure}\usepackage{epsfig}\usepackage{epstopdf}
\usepackage{graphics}\usepackage{subfigure}\usepackage{upgreek}\usepackage{caption}\usepackage{theorem}
\usepackage{breqn}
\usepackage{multicol}
\usepackage{eufrak}
\usepackage{eucal}
\usepackage{array}
\usepackage[compress]{cite}
\usepackage{algorithm}
\usepackage{algorithmic}
\allowdisplaybreaks[4]

\newtheorem{lemma}{Lemma}\newtheorem{proposition}{Proposition}\theoremheaderfont{\normalfont\bfseries}

\makeatother

\begin{document}
\title{User Cooperation for IRS-aided Secure SWIPT MIMO Systems}
\author{Gui~Zhou, Cunhua~Pan, Hong~Ren, Kezhi~Wang,  Kok Keong Chai, and Kai-Kit~Wong,~\IEEEmembership{Fellow, IEEE}
 \thanks{(Corresponding author: Cunhua Pan)
 	
 	 G. Zhou, C. Pan, and K.-K. Chai are with the
School of Electronic Engineering and Computer Science at Queen Mary
University of London, London E1 4NS, U.K. (e-mail: g.zhou, c.pan, michael.chai@qmul.ac.uk).

H. Ren is with the National Mobile Communications Research Laboratory,
Southeast University, Nanjing 210096, China. (hren@seu.edu.cn).

K. Wang is with Department of Computer and Information Sciences, Northumbria
University, UK. (e-mail: kezhi.wang@northumbria.ac.uk).

K.-K. Wong is with the Department of Electronic and Electrical Engineering,
University College London, London WC1E 6BT, U.K. (e-mail: kai-kit.wong@ucl.ac.uk).}}
\maketitle
\begin{abstract}
In this paper, intelligent reflecting surface (IRS) is proposed to
enhance the physical layer security in the Rician fading channel where
the angular direction of the eavesdropper is aligned with a legitimate
user. In this scenario, we consider a two-phase communication system
under the active attacks and passive eavesdropping. Particularly,
in the first phase, the base station avoids direct transmission to
the attacked user. While, in the second phase, other users cooperate
to forward signals to the attacked user with the help of IRS and energy
harvesting technology. Under the active attacks, we investigate an
outage constrained beamforming design problem under the statistical
cascaded channel error model, which is solved by using the Bernstein-type
inequality. As for the passive eavesdropping, an average secrecy rate
maximization problem is formulated, which is addressed by a low complexity
algorithm. Numerical results show that the negative effect of the
eavesdropper's channel error is greater than that of the legitimate
user. 
\end{abstract}

\begin{IEEEkeywords}
Intelligent reflecting surface (IRS), reconfigurable intelligent surface
(RIS), robust design, energy harvesting, physical layer security. 
\end{IEEEkeywords}

\section{Introduction}

Communication security is widely regarded as one of the important
issues in wireless communications. Traditionally, security is enforced
by imposing cryptographic protocols in the application layer \cite{cryptography1976}.
However, this upper layer solution is not flexible as it requires
complex key exchange protocols. Fortunately, it is shown by Wyner
\cite{Wyner1975} that secure communication can be guaranteed by adopting
advanced signal processing techniques developed in the physical layer.
In specific, these techniques exploit differences in channel conditions
and interference environment to enhance the received signal of legitime
users (LUs) and suppress the signal received by the eavesdropper (ED).

In order to enhance physical layer security, intelligent reflecting
surface (IRS), a kind of passive metasurface, has emerged as a promising
technique \cite{Marco-4,Marco-3,KK2020}. By changing the reflection
direction of the incident signal, the IRS is capable of reconfiguring
the wireless channels, enhancing the signal from the base station
(BS) to LU and suppressing the radio frequency (RF) power leaked to
the ED \cite{Shen2019secrecy,sheng2020,Guan-sec,yu-robust,Hong-robust}.
Furthermore, the IRS can be readily coated on existing buildings,
such as the walls and ceilings, which reduces the cost and complexity
of deployment operations. Hence, IRS holds great promise for security
enhancement as it provides a cost-effective and energy-efficient approach.

In general, ED works in two modes: active attacks and passive eavesdropping
\cite{PLS-1,PLS-2}. In an active attack, in order to mislead the
BS to send signals to the ED, the ED pretends to be a LU sending pilot
signals to the BS during the channel estimation procedure. Nonetheless,
a passive attack is more challenging to tackle since the passive ED
can hide itself and its channel state information (CSI) is not available
at the BS.

Recently, the benefits of IRS in physical layer security under the
active attacks have been investigated in the existing literature \cite{Shen2019secrecy,sheng2020,Guan-sec,yu-robust,Hong-robust}.
The performance gains of IRS in terms of security capacity was first
explored in a simple model consisting of only one single-antenna LU
and one single-antenna ED in \cite{Shen2019secrecy}. Closed-form
solutions of the phase shifters of IRS were obtained by leveraging
the majorization-minimization (MM) technique in \cite{Shen2019secrecy},
which had a better performance than the classical semidefinite relaxation
(SDR) method. The authors of \cite{sheng2020} extended the results
of \cite{Shen2019secrecy} to a multiple-input multiple-output (MIMO)
system where artificial noise (AN) was introduced to enhance the security
performance. The results of \cite{Guan-sec} further showed that the
AN-aided system without an IRS outperforms the IRS-aided system without
AN when the IRS is surrounded by a large number of eavesdroppers.
However, all the above contributions were based on the assumption
of perfect CSI of the eavesdropping channels at the BS. This assumption
is too strict and even impractical. The reasons are twofold: 1) It
is challenging to estimate the IRS-related channels since IRS is passive
and can neither send nor receive pilot signals. 2) The pilot transmission
from the ED to the BS may not be continuous and the corresponding
CSI at the BS may be outdated. To deal with the imperfect CSI of the
ED, robust transmission methods for secure communication of IRS were
proposed in \cite{yu-robust,Hong-robust}. In particular, the authors
of \cite{yu-robust} proposed a worst-case robust secure transmission
strategy under the assumption of imperfect CSI from the IRS to the
ED. On the other hand, the authors of \cite{Hong-robust} considered
the more practical imperfect cascaded BS-IRS-ED channel and proposed
an outage constrained beamforming design method under the statistical
CSI error model. However, the imperfect CSI of both LU and ED was
not studied in \cite{Hong-robust}.

To the best of our knowledge, all the existing contributions on the
IRS-aided security enhancement were developed under the active attacks,
where the BS can acquire the CSI of ED. There is no existing work
studying the passive eavesdropping in IRS-aided secure communication
systems. In addition, even for the imperfect CSI under the active
attacks, the methods proposed in \cite{yu-robust,Hong-robust} are
only applicable to small-size IRS (i.e., the number of the reflection
elements is less than 10) which can be observed from the numerical
simulations. The limitations for the research of small-size IRS lie
in twofold. The first is that some interesting observations can be
found in the robust design in an IRS-aided communication system only
when the number of reflection elements is large enough \cite{GuiTSProbust}.
Secondly, IRS has advantages over the conventional massive MIMO and
relay in terms of energy efficiency only when the number of IRS reflection
elements is large \cite{Gui2019IRS}.

Against the above background, this paper studies the IRS-aided secrecy
communication under the active attacks and passive eavesdropping.
The contributions of this paper are summarized as follows: 
\begin{itemize}
\item This paper proposes an IRS-aided two-phase secrecy communication scheme
for a scenario where the ED has a similar channel direction as a LU
in order to asquire high-quality eavesdropping information. In particular,
in the multicasting phase, the BS transmits signals to the LU with
low transmission power to reduce the information leakage to the ED.
In the user cooperation phase, other LUs forward the received signals
to the attacked LU with the assistance of IRS by using the energy
harvested in the previous phase. In addition, two models of ED are
considered in this work, i.e. active attack and passive eavesdropping. 
\item In the presence of statistical CSI error under the active attack,
we develop an outage constrainted beamforming design that maximizes
the secrecy rate subject to the unit-modulus constraint, the energy
harvesting constraint and the secrecy rate outage probability constraint.
Here, the outage probability constraint guarantees the maximum secrecy
rate of the system for secure communication under a predetermined
probability. By resorting to the Bernstein-Type inequality (BTI) and
convex approximations, the non-convexity of constraints is addressed.
Then, the active precoders and the passive reflection beamforming
are updated by using the proposed semidefinite programming (SDP) and
penalty convex-concave procedure (CCP) technique respectively in an
iterative manner. 
\item For the passive ED case with only partial CSI, we maximize an average
secrecy rate subject to the unit-modulus constraint of the reflection
beamforming and the energy harvesting constraint. To address the numerical
integration in the objective function, an angular secrecy model, which
is analytically non-convex, is proposed. A low-complexity algorithm
is proposed based on the MM-based alternate optimization (AO) framework,
where the precoders are updated by solving a convex optimization problem
and the reflection beamforming is updated in a closed-form solution
which is globally optimal. 
\item The numerical results demonstrate that the level of the cascaded CSI
error plays a vital role in the IRS-aided secure communication systems.
In particular, at low error of cascaded CSI, the secrecy rate increases
with the number of elements at the IRS due to the increased beamforming
gain. However, at large level of cascaded CSI error, the secrecy rate
decreases with the number of elements at the IRS due to the increased
channel estimation error. Hence, whether to enable the IRS for enhancing
the security capacity in the communication systems depends on the
level of the cascaded CSI error. In addition, the IRS can enhance
the average secrecy rate under the passive eavesdropping. 
\end{itemize}
\,\,\,\,\,\,\,The remainder of this paper is organized as follows.
Section II introduces the channel model and the system model. Outage
constrained robust design problem is formulated for the active eavesdropper
model in Section III. Section IV further investigates the average
eavesdropping rate maximization problem under the passive eavesdropping.
Finally, Section V and Section VI show the numerical results and conclusions,
respectively.

\noindent \textbf{Notations:} The following mathematical notations
and symbols are used throughout this paper. Vectors and matrices are
denoted by boldface lowercase letters and boldface uppercase letters,
respectively. The symbols $\mathbf{X}^{*}$, $\mathbf{X}^{\mathrm{T}}$,
$\mathbf{X}^{\mathrm{H}}$, and $||\mathbf{X}||_{F}$ denote the conjugate,
transpose, Hermitian (conjugate transpose), Frobenius norm of matrix
$\mathbf{X}$, respectively. The symbol $||\mathbf{x}||_{2}$ denotes
2-norm of vector $\mathbf{x}$. The symbols $\mathrm{Tr}\{\cdot\}$,
$\mathrm{Re}\{\cdot\}$, $|\cdot|$, $\lambda(\cdot)$, and $\angle\left(\cdot\right)$
denote the trace, real part, modulus, eigenvalue, and angle of a complex
number, respectively. $\mathrm{diag}(\mathbf{x})$ is a diagonal matrix
with the entries of $\mathbf{x}$ on its main diagonal. $[\mathbf{x}]_{m}$
means the $m^{\mathrm{th}}$ element of the vector $\mathbf{x}$.
The Kronecker product and the Hadamard product between two matrices
$\mathbf{X}$ and $\mathbf{Y}$ is denoted by $\mathbf{X}\otimes\mathbf{Y}$
and $\mathbf{X}\odot\mathbf{Y}$, respectively. $\mathbf{X}\succeq\mathbf{Y}$means
that $\mathbf{X}-\mathbf{Y}$ is positive semidefinite. Additionally,
the symbol $\mathbb{C}$ denotes complex field, $\mathbb{R}$ represents
real field, and $j\triangleq\sqrt{-1}$ is the imaginary unit.

\section{System Model}

\begin{figure}
\centering \includegraphics[width=3in,height=2in]{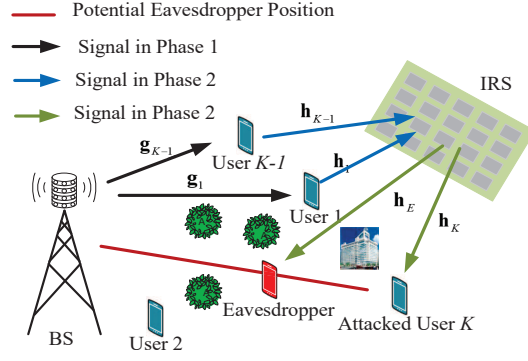}
\caption{Two-phase communication system}
\label{channel-model} 
\end{figure}

As shown in Fig. \ref{channel-model}, we consider Rician wiretap
channels where a BS with $N$ transmit antennas communicates with
$K$ single-antenna LUs in the presence of a single-antenna ED. An
IRS with $M$ reflection elements is introduced to aid the secure
communication.

\subsection{Channel Model}

\begin{figure}
\centering \includegraphics[width=3in,height=1.2in]{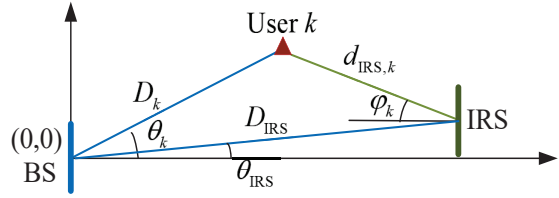}
\caption{Coordinates of communication nodes in the system}
\label{channel-model-1} 
\end{figure}

Define the set of all LUs as $\mathcal{K}=\{1,2,...,K\}$, and denote
set $\mathcal{K}_{-K}=\mathcal{K}/\{K\}$ and set $\mathcal{K}_{+E}=\mathcal{K}\cup\{E\}$.
By denoting $\{D_{i},\theta_{i}\}_{\forall i\in\mathcal{\mathcal{K}}_{+E}}$
as the distances and the azimuth angles respectively from the BS to
the LUs and the ED, as shown in Fig. \ref{channel-model}, then the
corresponding channels $\{\mathbf{g}_{i}\in\mathbb{C}^{N\times1}\}_{\forall i\in\mathcal{\mathcal{K}}_{+E}}$
follow the Rician fading distribution \cite{AAUC}: 
\begin{align}
\mathbf{g}_{i}  =\sqrt{\varrho_{0}\left(\frac{D_{i}}{d_{0}}\right)^{-\alpha_{\mathrm{BS}}}}\Biggl(\sqrt{\frac{K_{\mathrm{BS}}}{1+K_{\mathrm{BS}}}}\mathbf{g}_{i}^{\mathrm{LOS}}+\sqrt{\frac{1}{1+K_{\mathrm{BS}}}}\mathbf{g}_{i}^{\mathrm{NLOS}}\Biggr),\forall i\in\mathcal{K}_{+E},
\end{align}
where $\varrho_{0}$ is the pathloss at the reference distance of
$d_{0}$, $\alpha_{BS}$ and $K_{BS}$ are the pathloss exponent and
the Rician factor of the BS-related links, respectively. It is assumed
that the BS is equipped with a uniform linear array (ULA). Then, the
line-of-sight (LoS) component is given by $\mathbf{g}_{i}^{\mathrm{LOS}}=\left[1,e^{-\mathrm{j}\pi\sin\theta_{i}},\cdots,e^{-\mathrm{j}(N-1)\pi\sin\theta_{i}}\right]$,
and the non-LoS component is drawn from a Rayleigh fading, i.e., $\mathbf{g}_{i}^{\mathrm{NLOS}}\sim\mathcal{CN}(\mathbf{0},\mathbf{I}_{N})$.

Furthermore, by denoting $\{D_{\mathrm{IRS}},\theta_{\mathrm{IRS}}\}$
as the distance and the azimuth angle from the BS to the IRS, it is
straightforward to obtain the distances $\{d_{\mathrm{IRS},i}\}_{\forall i\in\mathcal{\mathcal{K}}_{+E}}$
and the azimuth angles $\{\varphi_{i}\}_{\forall i\in\mathcal{\mathcal{K}}_{+E}}$
from the IRS to the LUs and the ED as shown in Fig. \ref{channel-model-1},
i.e., 
\begin{align*}
d_{\mathrm{IRS},i} & =\Bigl[(D_{\mathrm{IRS}}\cos\theta_{\mathrm{IRS}}-D_{i}\cos\theta_{i})^{2}+(D_{\mathrm{IRS}}\sin\theta_{\mathrm{IRS}}-D_{i}\sin\theta_{i})^{2}\Bigr]^{-1/2},\\
\sin\varphi_{i} & =\frac{1}{d_{\mathrm{IRS},i}}(D_{i}\sin\theta_{i}-D_{\mathrm{IRS}}\sin\theta_{\mathrm{IRS}}),\\
\cos\varphi_{i} & =\frac{1}{d_{\mathrm{IRS},i}}(D_{\mathrm{IRS}}\cos\theta_{\mathrm{IRS}}-D_{i}\cos\theta_{i}).
\end{align*}
The corresponding channels $\{\mathbf{h}_{i}\in\mathbb{C}^{N\times1}\}_{\forall i\in\mathcal{\mathcal{K}}_{+E}}$
are given by 
\begin{align}
\mathbf{h}_{i} & =\sqrt{\varrho_{0}\left(\frac{d_{\mathrm{IRS},i}}{d_{0}}\right)^{-\alpha_{\mathrm{IRS}}}}\Biggl(\sqrt{\frac{K_{\mathrm{IRS}}}{1+K_{\mathrm{IRS}}}}\mathbf{h}_{i}^{\mathrm{LOS}}+\sqrt{\frac{1}{1+K_{\mathrm{IRS}}}}\mathbf{h}_{i}^{\mathrm{NLOS}}\Biggr),\forall i\in\mathcal{K}_{+E},\label{eq:los}
\end{align}
where $\alpha_{\mathrm{IRS}}$ and $K_{\mathrm{IRS}}$ are the pathloss
exponent and the Rician factor of the IRS-related links, respectively.
The non-LoS component follows the distribution of $\mathbf{h}_{i}^{\mathrm{NLOS}}\sim\mathcal{CN}(\mathbf{0},\mathbf{I}_{M})$.
It is assumed that the IRS is an uniform plane array (UPA) with size
of $M=M_{{\rm x}}M_{{\rm y}}$, where $M_{{\rm x}}$ and $M_{{\rm y}}$
are the number of reflection elements in x-axis and y-axis, respectively.
Then, the LoS component is written as 
\begin{align*}
\mathbf{h}_{i}^{\mathrm{LOS}}=  [1,\cdots,e^{-\mathrm{j}\pi(x\cos\varphi_{i}\cos\phi+y\sin\varphi_{i}\cos\phi)\sin\theta_{i}},\cdots, e^{-\mathrm{j}\pi((M_{{\rm x}}-1)\cos\varphi_{i}\cos\phi+(M_{y}-1)\sin\varphi_{i}\cos\phi)\sin\theta_{i}}],
\end{align*}
where $1\le x\le M_{{\rm x}}$, $1\le y\le M_{{\rm y}}$ and $\phi$
is the elevation angle observed at the IRS side.

\subsection{Signal Transmission}

To achieve high success rate of attack, the ED can locate the line
between the BS and legitimate user. In this situation, the signal
received by the ED is highly correlated with that of this user \cite{PLS-1,PLS-2},
thus posing great threats to the system. As shown in Fig. 1, we assume
that the ED hides at the line connecting the BS and one of the users,
denoted as user $K$, which leads to $\theta_{E}\approx\theta_{K}$,
$\mathbf{g}_{E}^{\mathrm{LOS}}\approx\mathbf{g}_{K}^{\mathrm{LOS}}$
and $D_{E}\in(0,D_{K})$. When the Rician factor $K_{\mathrm{BS}}$
is sufficiently large, the channel $\mathbf{g}_{E}$ is approximately
equal to the channel $\mathbf{g}_{K}$.

In order to achieve high-quality secure communication, the angle aware
user cooperation (AAUC) scheme \cite{AAUC} is adopted here. In particular,
in the first phase, the BS multicasts the common signal to all users
except user $K$. In the second phase, the helping users $(\forall k\in\mathcal{K}_{-K})$
forward the decoded common signal to user $K$ via the IRS. In this
work, in order to implement the AAUC scheme without consuming extra
energy, the LUs adopt the hybrid information and energy harvesting
receiving mode which splits the received signal into two power streams
with power splitting ratios $t_{k}$ and $1-t_{k}$. The former is
used for decoding the signal and the latter is for energy harvesting.

\subsubsection{Multicasting Phase}

In this phase, the BS multicasts the signal $s$ to the helping LUs
through beamforming vector ${\bf f}\in\mathbb{C}^{N\times1}$ which
is limited to the maximum transmit power $P_{\max}$, i.e., $||{\bf f}||_{2}^{2}\leq P_{\max}$.
Since $\mathbf{g}_{E}\approx\mathbf{g}_{K}$, the beamforming ${\bf f}$
needs to satisfy $|\mathbf{g}_{K}^{\mathrm{H}}{\bf f}|=0$ to ensure
that $|\mathbf{g}_{E}^{\mathrm{H}}{\bf f}|\approx0$. Let $\mathbf{Q}\in\mathbb{C}^{N\times\text{(}N-1)}$
be the orthogonal matrix which spans the null space of $\mathbf{g}_{K}$
by using the QR decomposition, i.e., $\mathbf{Q}^{\mathrm{H}}\mathbf{Q}={\bf I}$.
Then, we can design ${\bf f}=\mathbf{Q}\mathbf{z}$, where $\mathbf{z}\in\mathbb{C}^{\text{(}N-1)\times1}$
is a newly introduced variable. Therefore, the signal received by
LU $k$ is given by $\mathbf{g}_{k}^{\mathrm{H}}\mathbf{Q}\mathbf{z}+n_{k}$,
where $n_{k}$ is the received noise with the noise power of $\sigma_{k}^{2}$.
By adopting the hybrid receiving mode and let ${\bf {t}}=[t_{1},...,t_{K-1}]^{{\rm {T}}}$
where $t_{k}$ is the power splitting ratio of LU $k$, the achievable
rate at LU $k\neq K$ is 
\begin{equation}
R_{k}\left({\bf z},t_{k}\right)=\frac{1}{2}\log_{2}\left(1+\frac{t_{k}}{\sigma_{k}^{2}}\left|\mathbf{g}_{k}^{\mathrm{H}}\mathbf{Q}\mathbf{z}\right|^{2}\right),\label{eq:rate-k}
\end{equation}
where the factor 1/2 is due to the two transmission phases. The harvested
power at LU $k\neq K$ is 
\begin{equation}
(1-t_{k})\left|\mathbf{g}_{k}^{\mathrm{H}}\mathbf{Q}\mathbf{z}\right|^{2}.
\end{equation}

\subsubsection{User Cooperation Phase}

In this phase, the helping LUs $(\forall k\in\mathcal{K}_{-K})$ forward
the signal $s$ to LU $K$ through a beamforming vector ${\bf w}\in\mathbb{C}^{(K-1)\times1}=[w_{1},...,w_{K-1}]^{\mathrm{T}}$
by using the power harvested in the multicasting phase. Since LU $K$
is randomly selected by the ED and assume that there are many obstacles
in the communication environment, such as indoor applications, the
direct links between the helping LUs and the LU $K$ may be blocked.
To address this issue, an IRS can be installed on the building with
a certain height, and thus the IRS is capable of reflecting the signals
forwarded by the helping LUs to LU $K$. Denote by $\mathbf{e}$ the
reflection coefficient vector of the IRS, where $|e_{m}|^{2}=1,\forall m=1,\cdots,M$.
Then, the signal received by LU $K$ is given by 
\begin{align*}
y_{K} & =\mathbf{h}_{K}^{\mathrm{H}}\mathrm{diag}(\mathbf{e}^{*})\mathbf{H}_{\mathrm{IRS}}\mathbf{w}s+\sigma_{K}^{2}\\
 & =\mathbf{e}^{\mathrm{H}}\mathbf{H}_{K}\mathbf{w}s+\sigma_{K}^{2},
\end{align*}
where $\mathbf{H}_{\mathrm{IRS}}=\left[\mathbf{h}_{1}\text{,\thinspace\thinspace...,\thinspace\thinspace\ensuremath{\mathbf{h}_{K-1}}}\right]$,
$\mathbf{H}_{K}=\left[\mathbf{h}_{K}^{*}\odot\mathbf{h}_{1}\text{,\thinspace\thinspace...,\thinspace\thinspace\ensuremath{\mathbf{h}_{K}^{*}\odot\mathbf{h}_{K-1}}}\right]$
is the cascaded LU-IRS-LU (LIL) channel, and $n_{K}\sim\mathcal{CN}(\mathbf{0},\sigma_{K}^{2})$
is the noise. The corresponding achievable rate is 
\begin{equation}
R_{K}\left(\mathbf{w},\mathbf{e}\right)=\frac{1}{2}\log_{2}\left(1+\frac{1}{\sigma_{K}^{2}}\left|\mathbf{e}^{\mathrm{H}}\mathbf{H}_{K}\mathbf{w}\right|^{2}\right).\label{eq:rate-K}
\end{equation}
On the other hand, the signal recieved by the ED is $y_{E}=\mathbf{e}^{\mathrm{H}}\mathbf{H}_{E}\mathbf{w}s+n_{E}$,
where $\mathbf{H}_{E}=\left[\mathbf{h}_{E}^{*}\odot\mathbf{h}_{1}\text{,\thinspace\thinspace...,\thinspace\thinspace\ensuremath{\mathbf{h}_{E}^{*}\odot\mathbf{h}_{K-1}}}\right]$
is the cascaded LU-IRS-ED (LIE) channel, and $n_{E}\sim\mathcal{CN}(\mathbf{0},\sigma_{E}^{2})$
is the received noise at the ED.

The corresponding eavesdropping rate is 
\begin{equation}
R_{E}\left(\mathbf{w},\mathbf{e}\right)=\frac{1}{2}\log_{2}\left(1+\frac{1}{\sigma_{E}^{2}}\left|\mathbf{e}^{\mathrm{H}}\mathbf{H}_{E}\mathbf{w}\right|^{2}\right).\label{eq:rate-E}
\end{equation}

Finally, the secrecy rate of this system under the AAUC scheme can
be expressed as \cite{PLS-2}: 
\begin{equation}
\left[\min_{\forall k\in\mathcal{K}}R_{k}-R_{E}\right]^{+}.
\end{equation}

In the following two sections, we consider the system design for two
ED models: the active eavesdropper model and the passive eavesdropper
model.

\section{ED Model I: Active Eavesdropper Model}

In this section, we consider the active attack case, in which the
ED pretends to be an LU sending pilot signals to the transmitters
(including the BS and the helping LUs) during the channel estimation
procedure \cite{PLS-1,PLS-2}. It is reasonable to assume that the
BS is capable of addressing this attack by using the multi-antenna
technique, so as to obtain perfect CSI of the system. Nevertheless,
the signle-antenna helping LUs only have the imperfect CSI of LU $K$
and the ED due to their limited anti-interference ability.

\subsection{Channel Uncertainties}

Based on the above assumption, the cascaded channels can be modeled
as 
\begin{align}
\mathbf{H}_{K} & =\widehat{\mathbf{H}}_{K}+\boldsymbol{\bigtriangleup}_{K},\thinspace\thinspace\mathbf{H}_{E}=\widehat{\mathbf{H}}_{E}+\boldsymbol{\bigtriangleup}_{E},
\end{align}
where $\widehat{\mathbf{H}}_{K}$ and $\widehat{\mathbf{H}}_{E}$
are the estimated cascaded channels, $\boldsymbol{\bigtriangleup}_{K}=[\boldsymbol{\bigtriangleup}_{1}^{K}\cdots\boldsymbol{\bigtriangleup}_{K-1}^{K}]$
and $\boldsymbol{\bigtriangleup}_{E}=[\boldsymbol{\bigtriangleup}_{1}^{E}\cdots\boldsymbol{\bigtriangleup}_{K-1}^{E}]$
are the unknown cascaded channel errors. $\boldsymbol{\bigtriangleup}_{k}^{K}$
and $\boldsymbol{\bigtriangleup}_{k}^{E}$ are the unknown cascaded
LIL and LIE channel error vectors at LU $k$, respectively.

According to \cite{GuiTSProbust}, the robust beamforming under the
statistical CSI error model outperforms the bounded CSI error model
in terms of the minimum transmit power, convergence speed and computational
complexity. In addition, the statistical channel error model is more
suitable to model the channel estimation error when the channel estimation
is based on the minimum mean sum error (MMSE) method. Hence, we adopt
the statistical model to characterize the cascadepd CSI imperfection
\cite{GuiTSProbust}, i.e., each CSI error vector is assumed to follow
the circularly symmetric complex Gaussian (CSCG) distribution, i.e.,
\begin{subequations}\label{Pro:statistic-error} 
\begin{align}
\boldsymbol{\bigtriangleup}_{k}^{K} & \sim\mathcal{CN}(\mathbf{0},\boldsymbol{\Sigma}_{k}^{K}),\boldsymbol{\Sigma}_{k}^{K}\succeq\mathbf{0},\forall k\in\mathcal{K}_{-K},\\
\boldsymbol{\bigtriangleup}_{k}^{E} & \sim\mathcal{CN}(\mathbf{0},\boldsymbol{\Sigma}_{k}^{E}),\boldsymbol{\Sigma}_{k}^{E}\succeq\mathbf{0},\forall k\in\mathcal{K}_{-K},
\end{align}
\end{subequations}where $\boldsymbol{\Sigma}_{k}^{K}\in\mathbb{C}^{M\times M}$
and $\boldsymbol{\Sigma}_{k}^{E}\in\mathbb{C}^{M\times M}$ are positive
semidefinite error covariance matrices. Note that the CSI error vectors
of different LUs are independent with each other. Therefore, we have
\begin{equation}
\mathrm{vec}(\boldsymbol{\bigtriangleup}_{K})\sim\mathcal{CN}(\mathbf{0},\boldsymbol{\Sigma}_{K}),\thinspace\thinspace\mathrm{vec}(\boldsymbol{\bigtriangleup}_{E})\sim\mathcal{CN}(\mathbf{0},\boldsymbol{\Sigma}_{E}),\label{eq:error}
\end{equation}
where $\boldsymbol{\Sigma}_{K}$ and $\boldsymbol{\Sigma}_{E}$ are
block diagonal matrices, i.e., $\boldsymbol{\Sigma}_{K}=\mathrm{diag}(\boldsymbol{\Sigma}_{1}^{K},...,\boldsymbol{\Sigma}_{K-1}^{K})$
and $\boldsymbol{\Sigma}_{E}=\mathrm{diag}(\boldsymbol{\Sigma}_{1}^{E},...,\boldsymbol{\Sigma}_{K-1}^{E})$.

\subsection{Outage constrained beamforming design}

\label{Pro:act-solution}

Under the statistical CSI error model, we develop a probabilistically
robust algorithm for the secrecy rate maximization problem, which
is formulated as\begin{subequations}\label{pro:outage-1} 
\begin{align}
\max\limits _{R_{\mathrm{sec}},\mathbf{z},\mathbf{w},\mathbf{e},\mathbf{t}} & R_{\mathrm{sec}}\\
\textrm{s.t.} & \mathrm{Pr}\left\{ \min_{\forall k\in\mathcal{K}}R_{k}-R_{E}\geq R_{\mathrm{sec}}\right\} \geq1-\rho\label{eq:r-c}\\
 & ||{\bf z}||_{2}^{2}\leq P_{\max}\label{eq:p-c}\\
 & \;|e_{m}|^{2}=1,1\leq m\leq M\label{eq:e-c}\\
 & 0\leq\mathbf{t}\leq1\label{eq:sp-c}\\
 & |w_{k}|^{2}\leq(1-t_{k})\left|\mathbf{g}_{k}^{\mathrm{H}}\mathbf{Q}\mathbf{z}\right|^{2},\forall k\in\mathcal{K}_{-K},\label{eq:ph-c}
\end{align}
\end{subequations}where $\rho\in(0,1]$ is the secrecy rate outage
probability.

Problem (\ref{pro:outage-1}) is difficult to solve due to the computationally
intractable rate outage probability constraint (\ref{eq:r-c}), the
non-convex unit-modulus constraint (\ref{eq:e-c}), and the non-convex
power constraint (\ref{eq:ph-c}).

Firstly, we replace constraint (\ref{eq:r-c}) with the development
of a safe approximation consisting of three steps in the following.

\textit{Step 1: Decouple the Probabilistic Constraint:} First of all,
based on the independence between $\{\mathbf{g}_{k}\}_{\forall k\in\mathcal{K}_{-K}}$
and $\mathbf{H}_{K}$, we have 
\begin{align}
(\ref{eq:r-c})\Leftrightarrow & \prod_{k=1}^{K}\mathrm{Pr}\left\{ R_{k}-R_{E}\geq R_{\mathrm{sec}}\right\} \geq1-\rho\label{eq:p-2}\\
\Leftarrow & \mathrm{Pr}\left\{ R_{k}-R_{E}\geq R_{\mathrm{sec}}\right\} \geq1-\bar{\rho},\forall k\in\mathcal{K}_{K},\label{eq:p-1}
\end{align}
where $\bar{\rho}=1-(1-\rho)^{1/K}$.

\textit{Step 2: Convenient Approximations: }To address the non-concavity
of $R_{k}-R_{E},\forall k\in\mathcal{K}_{K}$, we need to construct
a sequence of surrogate functions of $\{R_{i}\}_{\forall i\in\mathcal{K}_{+E}}$.
More specifically, we need the following lemmas.

\begin{lemma}\label{lower-1} \cite{book-convex} The quadratic-over-linear
function $\frac{x^{2}}{y}$ is jointly convex in $(x,y)$, and lower
bounded by its linear first-order Tayler approximation $\frac{2x^{(n)}}{y^{(n)}}x-(\frac{x^{(n)}}{y^{(n)}})^{2}y$
at fixed point $(x^{(n)},y^{(n)})$.

\end{lemma} By substituting $x$ with $\mathbf{g}_{k}^{\mathrm{H}}\mathbf{Q}\mathbf{z}$
and $y$ with $1/t_{k}$, we utilize Lemma \ref{lower-1} to obtain
a concave lower bound of rate $R_{k}\left(\mathbf{z},t_{k}\right)$
for $\forall k\in\mathcal{K}_{-K}$. The lower bound is given by 
\begin{align}
\widetilde{R}_{k}(\mathbf{z},t_{k}|\mathbf{z}^{(n)},  t_{k}^{(n)})=\frac{1}{2}\log_{2}\Biggl(1-\frac{t_{k}^{(n),2}}{\sigma_{k}^{2}t_{k}}\left|\mathbf{g}_{k}^{\mathrm{H}}\mathbf{Q}\mathbf{z}^{(n)}\right|^{2} +2t_{k}^{(n)}\mathrm{Re}\left\{ \frac{1}{\sigma_{k}^{2}}{\bf z}^{(n),\mathrm{H}}\mathbf{Q}^{\mathrm{H}}\mathbf{g}_{k}\mathbf{g}_{k}^{\mathrm{H}}\mathbf{Q}\mathbf{z}\right\} \Biggr)\label{eq:c}
\end{align}
for any feasible solution $\{\mathbf{z}^{(n)},t_{k}^{(n)}\}$.

\begin{lemma}\label{upper}

The upper bound of rate $R_{E}\left(\mathbf{w},\mathbf{e}\right)$
is given by 
\begin{align*}
\widetilde{R}_{E}\left(\mathbf{w},\mathbf{e},a_{E}\right) & =\frac{a_{E}\left|\mathbf{e}^{\mathrm{H}}\mathbf{H}_{E}\mathbf{w}\right|^{2}/\sigma_{E}^{2}+a_{E}-\ln a_{E}-1}{2\ln2},
\end{align*}
where $a_{E}$ is the auxiliary variable.

\end{lemma}

\textbf{\textit{Proof: }}See Appendix \ref{subsec:The-proof-of-1}.\hspace{4.5cm}$\blacksquare$

\begin{lemma}\label{lower-2}The lower bound of rate $R_{K}\left(\mathbf{w},\mathbf{e}\right)$
is given by 
\begin{align*}
\widetilde{R}_{K}(\mathbf{w},\mathbf{e},  a_{K},v)=\frac{1}{2\ln2}(-a_{K}|v|^{2}|\mathbf{e}^{\mathrm{H}}\mathbf{H}_{K}\mathbf{w}|^{2}-\sigma_{K}^{2}a_{K}|v|^{2} +2a_{K}\text{Re}\left\{ v\mathbf{e}^{\mathrm{H}}\mathbf{H}_{K}\mathbf{w}\right\} -a_{K}+\ln a_{K}+1),
\end{align*}
where $a_{K}$ and $v$ are the auxiliary variables.

\end{lemma}

\textbf{\textit{Proof: }}See Appendix \ref{subsec:The-proof-of-2}.\hspace{4.5cm}$\blacksquare$

For the convenience of derivations, we assume that $\boldsymbol{\Sigma}_{k}^{K}=\varepsilon_{K,k}^{2}\mathbf{I}_{M}$
and $\boldsymbol{\Sigma}_{k}^{E}=\varepsilon_{E,k}^{2}\mathbf{I}_{M}$,
then $\boldsymbol{\Sigma}_{K}=\boldsymbol{\Lambda}_{K}\otimes\mathbf{I}_{M}$
where $\boldsymbol{\Lambda}_{K}=\mathrm{diag}(\varepsilon_{K,1}^{2},...,\varepsilon_{K,K-1}^{2})$,
and $\boldsymbol{\Sigma}_{E}=\boldsymbol{\Lambda}_{E}\otimes\mathbf{I}_{M}$
where $\boldsymbol{\Lambda}_{E}=\mathrm{diag}(\varepsilon_{E,1}^{2},...,\varepsilon_{E,K-1}^{2})$.
Furthermore, the error vectors in (\ref{eq:error}) can be rewritten
as $\mathrm{vec}(\boldsymbol{\bigtriangleup}_{K})=\boldsymbol{\Sigma}_{K}^{\frac{1}{2}}\mathbf{i}_{K}$
where $\mathbf{i}_{K}\sim\mathcal{CN}(\mathbf{0},\mathbf{I}_{M(K-1)})$,
and $\mathrm{vec}(\boldsymbol{\bigtriangleup}_{E})=\boldsymbol{\Sigma}_{E}^{\frac{1}{2}}\mathbf{i}_{E}$
where $\mathbf{i}_{E}\sim\mathcal{CN}(\mathbf{0},\mathbf{I}_{M(K-1)})$.
Define $\mathbf{E}=\mathbf{e}\mathbf{e}^{\mathrm{H}}$ and $\mathbf{W}=\mathbf{w}\mathbf{w}^{\mathrm{H}}$.
Combining (\ref{eq:c}) with Lemma \ref{upper}, the secrecy rate
outage probabilities for $\forall k\in\mathcal{K}_{-K}$ in (\ref{eq:p-1})
are equivalent to 
\begin{align}
 & \mathrm{Pr}\left\{ R_{k}-R_{E}\geq R_{\mathrm{sec}}\right\} \nonumber \\
\geq & \mathrm{Pr}\left\{ \widetilde{R}_{k}-\widetilde{R}_{E}\geq R_{\mathrm{sec}}\right\} \nonumber \\
= & \mathrm{Pr}\Bigl\{ a_{E}\mathrm{Tr}\left(\mathbf{E}(\widehat{\mathbf{H}}_{E}+\boldsymbol{\bigtriangleup}_{E})\mathbf{W}(\widehat{\mathbf{H}}_{E}^{\mathrm{H}}+\boldsymbol{\bigtriangleup}_{E}^{\mathrm{H}})\right)-[2\ln2(\widetilde{R}_{k}-R_{\mathrm{sec}})-a_{E}+\ln a_{E}+1]\sigma_{E}^{2}\leq0\Bigr\}\nonumber \\
= & \mathrm{Pr}\left\{ \mathbf{i}_{E}^{\mathrm{H}}\mathbf{U}_{E}\mathbf{i}_{E}+2\mathrm{Re}\left\{ \mathbf{u}_{E}^{\mathrm{H}}\mathbf{i}_{E}\right\} +u_{k}\leq0\right\} ,\label{eq:pro-k}
\end{align}
where\begin{subequations}\label{eq:Outage-K-2} 
\begin{align}
\mathbf{U}_{E} & =a_{E}\boldsymbol{\Sigma}_{E}^{\frac{1}{2}}(\mathbf{W}^{\mathrm{T}}\otimes\mathbf{E})\boldsymbol{\Sigma}_{E}^{\frac{1}{2}},\\
\mathbf{u}_{E} & =a_{E}\boldsymbol{\Sigma}_{E}^{\frac{1}{2}}\mathrm{vec}(\mathbf{E}\mathbf{\widehat{H}}_{E}\mathbf{W}),\\
u_{k} & =a_{E}\mathrm{Tr}\left(\mathbf{E}\mathbf{\widehat{H}}_{E}\mathbf{W}\widehat{\mathbf{H}}_{E}^{\mathrm{H}}\right)-[(\widetilde{R}_{k}-R_{\mathrm{sec}})2\ln2-a_{E}+\ln a_{E}+1]\sigma_{E}^{2}.
\end{align}
\end{subequations}

Combining Lemma \ref{upper} with Lemma \ref{lower-2}, the secrecy
rate outage probability for LU $K$ in (\ref{eq:p-1}) is equivalent
to 
\begin{align}
 & \mathrm{Pr}\left\{ R_{K}-R_{E}\geq R_{\mathrm{sec}}\right\} \nonumber \\
\geq & \mathrm{Pr}\left\{ \widetilde{R}_{K}-\widetilde{R}_{E}\geq R_{\mathrm{sec}}\right\} \nonumber \\
= & \mathrm{Pr}\Bigl\{ a_{K}|v|^{2}\mathrm{Tr}\left(\mathbf{E}(\widehat{\mathbf{H}}_{K}+\boldsymbol{\bigtriangleup}_{K})\mathbf{W}(\widehat{\mathbf{H}}_{K}^{\mathrm{H}}+\boldsymbol{\bigtriangleup}_{K}^{\mathrm{H}})\right) -2a_{K}\mathrm{Re}\left\{ v\mathbf{e}^{\mathrm{H}}(\widehat{\mathbf{H}}_{K}+\boldsymbol{\bigtriangleup}_{K})\mathbf{w}\right\} +\nonumber \\
 & \frac{a_{E}}{\sigma_{E}^{2}}\mathrm{Tr}\left(\mathbf{E}(\widehat{\mathbf{H}}_{E}+\boldsymbol{\bigtriangleup}_{E})\mathbf{W}(\widehat{\mathbf{H}}_{E}^{\mathrm{H}}+\boldsymbol{\bigtriangleup}_{E}^{\mathrm{H}})\right)-c\leq0\Bigr\}\nonumber \\
= & \mathrm{Pr}\left\{ \mathbf{i}^{\mathrm{H}}\mathbf{U}_{K}\mathbf{i}+2\mathrm{Re}\left\{ \mathbf{u}_{K}^{\mathrm{H}}\mathbf{i}\right\} +u_{K}\leq0\right\} ,\label{eq:pro-K}
\end{align}
where\begin{subequations}\label{eq:Outage-K-1} 
\begin{align}
\mathbf{i} & =[\begin{array}{cc}
\mathbf{i}_{K}^{\mathrm{H}}, & \mathbf{i}_{E}^{\mathrm{H}}\end{array}]^{\mathrm{H}},\\
\mathbf{U}_{K} & =\mathrm{diag}\Bigl\{ a_{K}|v|^{2}\boldsymbol{\Sigma}_{K}^{\frac{1}{2}}(\mathbf{W}^{\mathrm{T}}\otimes\mathbf{E})\boldsymbol{\Sigma}_{K}^{\frac{1}{2}},\ \frac{a_{E}}{\sigma_{E}^{2}}\boldsymbol{\Sigma}_{E}^{\frac{1}{2}}(\mathbf{W}^{\mathrm{T}}\otimes\mathbf{E})\boldsymbol{\Sigma}_{E}^{\frac{1}{2}}\Bigr\},\\
\mathbf{u}_{K} & =[a_{K}|v|^{2}\mathrm{vec}^{\mathrm{H}}(\mathbf{E}\mathbf{\widehat{H}}_{K}\mathbf{W})\boldsymbol{\Sigma}_{K}^{\frac{1}{2}}-a_{K}v\mathrm{vec}^{\mathrm{H}}(\mathbf{e}\mathbf{w}^{\mathrm{H}})\boldsymbol{\Sigma}_{K}^{\frac{1}{2}}, \ \frac{a_{E}}{\sigma_{E}^{2}}\mathrm{vec}^{\mathrm{H}}(\mathbf{E}\mathbf{\widehat{H}}_{E}\mathbf{W})\boldsymbol{\Sigma}_{E}^{\frac{1}{2}}]^{\mathrm{H}},\\
u_{K} & =a_{K}|v|^{2}\mathrm{Tr}\left(\mathbf{E}\mathbf{\widehat{H}}_{K}\mathbf{W}\widehat{\mathbf{H}}_{K}^{\mathrm{H}}\right)+\frac{a_{E}}{\sigma_{E}^{2}}\mathrm{Tr}\left(\mathbf{E}\mathbf{\widehat{H}}_{E}\mathbf{W}\widehat{\mathbf{H}}_{E}^{\mathrm{H}}\right) \ -2a_{K}\mathrm{Re}\left\{ v\mathbf{e}^{\mathrm{H}}\mathbf{\widehat{H}}_{K}\mathbf{w}\right\} -c,\\
c & =\ln a_{E}+\ln a_{K}-a_{E}-a_{K}-2R_{\mathrm{sec}}\ln2 \ -\sigma_{K}^{2}a_{K}|v|^{2}+2.
\end{align}
\end{subequations}

Now, substituting (\ref{eq:pro-k}) and (\ref{eq:pro-K}) into (\ref{eq:p-1}),
then (\ref{eq:p-1}) is replaced by\begin{subequations}\label{eq:out-1}
\begin{align}
 & \mathrm{Pr}\left\{ \mathbf{i}_{E}^{\mathrm{H}}\mathbf{U}_{E}\mathbf{i}_{E}+2\mathrm{Re}\left\{ \mathbf{u}_{E}^{\mathrm{H}}\mathbf{i}_{E}\right\} +u_{k}\leq0\right\} \geq1-\bar{\rho}, \forall k\in\mathcal{K}_{-K},\label{eq:OUT-k}\\
 & \mathrm{Pr}\left\{ \mathbf{i}^{\mathrm{H}}\mathbf{U}_{K}\mathbf{i}+2\mathrm{Re}\left\{ \mathbf{u}_{K}^{\mathrm{H}}\mathbf{i}\right\} +u_{K}\leq0\right\} \geq1-\bar{\rho}.\label{eq:OUT-K}
\end{align}
\end{subequations}

\textit{Step 3: A Bernstein-Type Inequality-Based Safe Approximation:
}The outage probabilities in (\ref{eq:out-1}) are characterized by
quadratic inequalities, which can be safely approximated by using
the following lemma.

\begin{lemma}\label{Bernstein-Type Inequalities-1} (Bernstein-Type
Inequality) \cite{BTI-II} Assume $f(\mathbf{x})=\mathbf{x}^{\mathrm{H}}\mathbf{U}\mathbf{x}+2\mathrm{Re}\{\mathbf{u}^{\mathrm{H}}\mathbf{x}\}+u$,
where $\mathbf{U}\in\mathbb{H}^{n\times n}$, $\mathbf{u}\in\mathbb{C}^{n\times1}$,
$u\in\mathbb{R}$ and $\mathbf{x}\in\mathbb{C}^{n\times1}\sim\mathcal{CN}(\mathbf{0},\mathbf{I})$.
Then for any $\rho\in[0,1]$, the following approximation holds: 
\begin{align}
 & \mathrm{Pr}\{\mathbf{x}^{\mathrm{H}}\mathbf{U}\mathbf{x}+2\mathrm{Re}\{\mathbf{u}^{\mathrm{H}}\mathbf{x}\}+u\leq0\}\geq1-\rho\nonumber \\
\Rightarrow & \mathrm{Tr}\left\{ \mathbf{U}\right\} +\sqrt{2\ln(1/\rho)}x-\ln(\rho)\lambda_{\max}^{+}(\mathbf{U})+u\leq0\nonumber \\
\Rightarrow & \left\{ \begin{array}{c}
\mathrm{Tr}\left\{ \mathbf{U}\right\} +\sqrt{2\ln(1/\rho)}x-\ln(\rho)y+u\leq0\\
\sqrt{||\mathbf{U}||_{F}^{2}+2||\mathbf{u}||_{2}^{2}}\leq x\\
y\mathbf{I}-\mathbf{U}\succeq\mathbf{0},y\geq0,
\end{array}\right.\label{eq:Outage-5-1}
\end{align}
where $\lambda_{\max}^{+}(\mathbf{U})=\max(\lambda_{\max}(\mathbf{U}),0)$.
$x$ and $y$ are slack variables. \hspace{6.5cm}$\blacksquare$

\end{lemma}

Before using Lemma \ref{Bernstein-Type Inequalities-1}, we need the
following simplified derivations for LU $k$, $\forall k\in\mathcal{K}_{-K}$,
i.e.,\begin{subequations}\label{eq:Outage-k} 
\begin{align}
 & \mathrm{Tr}\left\{ \mathbf{U}_{E}\right\} =\mathrm{Tr}\left\{ a_{E}\boldsymbol{\Sigma}_{E}^{\frac{1}{2}}(\mathbf{W}^{\mathrm{T}}\otimes\mathbf{E})\boldsymbol{\Sigma}_{E}^{\frac{1}{2}}\right\} \nonumber \\
 & \ \ \ \ \ \ \ \ \ \ =\mathrm{Tr}\left\{ a_{E}(\mathbf{W}^{\mathrm{T}}\otimes\mathbf{E})(\boldsymbol{\Lambda}_{E}\otimes\mathbf{I}_{M})\right\} \nonumber \\
 & \ \ \ \ \ \ \ \ \ \ =a_{E}M\mathrm{Tr}\left\{ \boldsymbol{\Lambda}_{E}\mathbf{W}\right\} ,\\
 & ||\mathbf{U}_{E}||_{F}^{2}=a_{E}^{2}M^{2}||\boldsymbol{\Lambda}_{E}\mathbf{W}||_{F}^{2},\\
 & ||\mathbf{u}_{E}||_{2}^{2}=a_{E}^{2}\mathrm{vec}^{\mathrm{H}}(\mathbf{E}\mathbf{\widehat{H}}_{E}\mathbf{W})(\boldsymbol{\Lambda}_{E}\otimes\mathbf{I}_{M})\mathrm{vec}(\mathbf{E}\mathbf{\widehat{H}}_{E}\mathbf{W})\nonumber \\
 & \ \ \ \ \ \ \ =a_{E}^{2}M||\boldsymbol{\Lambda}_{E}^{\frac{1}{2}}\mathbf{W}\mathbf{\widehat{H}}_{E}^{\mathrm{H}}\mathbf{e}||_{2}^{2},\\
 & \lambda_{\max}(\mathbf{U}_{E})=\lambda_{\max}(a_{E}\boldsymbol{\Sigma}_{E}^{\frac{1}{2}}(\mathbf{W}^{\mathrm{T}}\otimes\mathbf{E})\boldsymbol{\Sigma}_{E}^{\frac{1}{2}})\nonumber \\
 & \ \ \ \ \ \ \ \ \ \ \ \ =\lambda_{\max}(a_{E}(\boldsymbol{\Lambda}_{E}\mathbf{W}^{\mathrm{T}}\otimes\mathbf{E}))\nonumber \\
 & \ \ \ \ \ \ \ \ \ \ \ \ =a_{E}\lambda_{\max}(\boldsymbol{\Lambda}_{E}\mathbf{W})\lambda(\mathbf{E})=a_{E}M\mathbf{w}^{\mathrm{H}}\boldsymbol{\Lambda}_{E}\mathbf{w}.
\end{align}
\end{subequations}By substituting (\ref{eq:Outage-k}) into (\ref{eq:Outage-5-1})
and introducing slack variables $\{x_{E},y_{E}\}$, the constraints
for $\forall k\in\mathcal{K}_{-K}$ in (\ref{eq:OUT-k}) are transformed
into the following deterministic form: 
\begin{align}
 & \textrm{BTI}_{1}\triangleq\left\{ \begin{array}{c}
\mathrm{Tr}\left\{ \mathbf{U}_{E}\right\} +\sqrt{2\ln(1/\bar{\rho})}x_{E}-\ln(\bar{\rho})y_{E}\\
+u_{k}\leq0,\forall k\in\mathcal{K}_{-K}\\
\left\Vert \begin{array}{c}
a_{E}M\mathrm{vec}(\boldsymbol{\Lambda}_{E}\mathbf{W})\\
\sqrt{2M}a_{E}\boldsymbol{\Lambda}_{E}^{\frac{1}{2}}\mathbf{W}\mathbf{\widehat{H}}_{E}^{\mathrm{H}}\mathbf{e}
\end{array}\right\Vert \leq x_{E}\\
y_{E}\geq a_{E}M\mathbf{w}^{\mathrm{H}}\boldsymbol{\Lambda}_{E}\mathbf{w}.
\end{array}\right.\label{eq:BTI-1}
\end{align}

On the other hand, the simplified derivations for LU $K$ are given
by\begin{subequations}\label{eq:Outage-K} 
\begin{align}
 & \mathrm{Tr}\left\{ \mathbf{U}_{K}\right\} =a_{K}|v|^{2}M\mathrm{Tr}\left\{ \boldsymbol{\Lambda}_{K}\mathbf{W}\right\} +\frac{a_{E}}{\sigma_{E}^{2}}M\mathrm{Tr}\left\{ \boldsymbol{\Lambda}_{E}\mathbf{W}\right\} ,\label{eq:h1}\\
 & ||\mathbf{U}_{K}||_{F}^{2}=a_{K}^{2}|v|^{4}M^{2}||\boldsymbol{\Lambda}_{K}\mathbf{W}||_{F}^{2}+\frac{a_{E}^{2}}{\sigma_{E}^{4}}M^{2}||\boldsymbol{\Lambda}_{E}\mathbf{W}||_{F}^{2},\label{eq:h2}\\
 & ||\mathbf{u}_{K}||^{2}=M||\boldsymbol{\Lambda}_{K}^{\frac{1}{2}}\left(a_{K}|v|^{2}\mathbf{W}\mathbf{\widehat{H}}_{K}^{\mathrm{H}}\mathbf{e}-a_{K}v\mathbf{w}\right)||_{2}^{2}+\frac{a_{E}^{2}}{\sigma_{E}^{4}}M||\mathbf{e}^{\mathrm{H}}\mathbf{\widehat{H}}_{E}\mathbf{W}\boldsymbol{\Lambda}_{E}^{\frac{1}{2}}||_{2}^{2},\label{eq:h3}\\
 & \lambda_{\max}(\mathbf{U}_{K})=\max\left\{ a_{K}|v|^{2}M\mathbf{w}^{\mathrm{H}}\boldsymbol{\Lambda}_{K}\mathbf{w},\thinspace\thinspace\thinspace\frac{a_{E}}{\sigma_{E}^{2}}M\mathbf{w}^{\mathrm{H}}\boldsymbol{\Lambda}_{E}\mathbf{w}\right\} .\nonumber 
\end{align}
\end{subequations}By substituting the above equations into (\ref{eq:Outage-5-1})
and introducing slack variables $\{x_{K},y_{K}\}$, the constraint
for LU $K$ in (\ref{eq:OUT-K}) is transformed into the following
deterministic form: 
\begin{align}
  \textrm{BTI}_{2}\triangleq \left\{ \begin{array}{c}
\mathrm{Tr}\left\{ \mathbf{U}_{K}\right\} +\sqrt{2\ln(1/\bar{\rho})}x_{K}-\ln(\bar{\rho})y_{K}+u_{K}\leq0\\
\left\Vert \begin{array}{c}
a_{K}|v|^{2}M\mathrm{vec}(\boldsymbol{\Lambda}_{K}\mathbf{W})\\
a_{E}M\mathrm{vec}(\boldsymbol{\Lambda}_{E}\mathbf{W})/\sigma_{E}^{2}\\
\sqrt{2M}\boldsymbol{\Lambda}_{K}^{\frac{1}{2}}\left(a_{K}|v|^{2}\mathbf{W}\mathbf{\widehat{H}}_{K}^{\mathrm{H}}\mathbf{e}-a_{K}v\mathbf{w}\right)\\
\sqrt{2M}a_{E}\boldsymbol{\Lambda}_{E}^{\frac{1}{2}}\mathbf{W}\mathbf{\widehat{H}}_{E}^{\mathrm{H}}\mathbf{e}/\sigma_{E}^{2}
\end{array}\right\Vert \leq x_{K}\\
y_{K}\geq\lambda_{\max}(\mathbf{U}_{K}),y_{K}\geq0.
\end{array}\right.\label{eq:BTI-2}
\end{align}

Then, to handle the non-convex power constraint (\ref{eq:ph-c}),
we replace the right hand side of (\ref{eq:ph-c}) with its linear
lower bound 
\begin{align}
\Xi({\bf z},t_{k})= & 2(1-t_{k}^{(n)})\mathrm{Re}\left\{ {\bf z}^{(n),\mathrm{H}}\mathbf{Q}^{\mathrm{H}}\mathbf{g}_{k}\mathbf{g}_{k}^{\mathrm{H}}\mathbf{Q}\mathbf{z}\right\} -\frac{(1-t_{k}^{(n)})^{2}\left|\mathbf{g}_{k}^{\mathrm{H}}\mathbf{Q}\mathbf{z}^{(n)}\right|^{2}}{(1-t_{k})}\label{eq:ph-2}
\end{align}
at feasible point $\{\mathbf{z}^{(n)},t_{k}^{(n)}\}$ by adopting
the same first-order Taylor approximation used in Lemma \ref{lower-1}.

Therefore, based on (\ref{eq:BTI-1}), (\ref{eq:BTI-2}) and (\ref{eq:ph-2})
and denoting $\mathbf{x}=[x_{E},x_{K}]^{\mathrm{T}}$and $\mathbf{y}=[y_{E},y_{K}]^{\mathrm{T}}$,
the approximation problem of Problem (\ref{pro:outage-1}) is given
by\begin{subequations}\label{Pro:act-pro} 
\begin{align}
\max\limits _{R_{\mathrm{sec}},\mathbf{z},\mathbf{w},\mathbf{e},\mathbf{t},a_{K},a_{E},v,\mathbf{x},\mathbf{y}} & R_{\mathrm{sec}}\\
\textrm{s.t.} & (\ref{eq:BTI-1}),(\ref{eq:BTI-2}),(\ref{eq:p-c})-(\ref{eq:sp-c}),\\
 & |w_{k}|^{2}\leq\Xi({\bf z},t_{k}),\forall k\in\mathcal{K}_{-K}.\label{eq:ph2}
\end{align}
\end{subequations}

For given $\{\mathbf{e},a_{K},a_{E},v\}$, we introduce a new variable
$\mathbf{W}=\mathbf{w}\mathbf{w}^{\mathrm{H}}$ with $\mathrm{rank}(\mathbf{W})=1$.
However, different from the general semidefinite programming (SDP),
$\mathbf{w}$ and $\mathbf{W}$, here, coexist in (\ref{eq:v}) and
(\ref{eq:h3}). Therefore, the semidefinite relaxation (SDR) technique
is not applicable here. In order to handle this problem, we assume
$\mathbf{w}$ and $\mathbf{W}$ are two different variables. If $\mathrm{Tr}\left\{ \mathbf{W}\right\} =\lambda_{\max}(\mathbf{W})$,
then we have $\mathrm{rank}(\mathbf{W})=1$. If the obtained $\mathbf{W}$
is not rank one, we will have $\mathrm{Tr}\left\{ \mathbf{W}\right\} -\lambda_{\max}(\mathbf{W})>0$.
Therefore, we constrain $\mathrm{Tr}\left\{ \mathbf{W}\right\} -\lambda_{\max}(\mathbf{W})$
less than a very small real positive number threshold $\varepsilon$
to guarantee the rank-1 condition of $\mathbf{W}$, yielding the surrogate
constraint of rank-1 constraint as\begin{subequations}\label{Pro:W-c-3}
\begin{align}
 & \mathrm{Tr}\left\{ \mathbf{W}\right\} -\lambda_{\max}(\mathbf{W})\leq\varepsilon.\label{eq:rank1}
\end{align}
\end{subequations}When $\mathrm{rank}(\mathbf{W})\approx1$, the
relationship between $\mathbf{w}$ and $\mathbf{W}$ is given by the
following constraint:\begin{subequations}\label{Pro:W-c} 
\begin{align}
 & -\varepsilon\leq||\mathbf{w}||^{2}-\mathrm{Tr}\left\{ \mathbf{W}\right\} \leq\varepsilon.\label{eq:rank2}
\end{align}
\end{subequations}

As for constraint (\ref{eq:rank1}), since $\lambda_{\max}(\mathbf{W})$
is a convex function of $\mathbf{W}$ \cite{book-convex}, the left
hand side of (\ref{eq:rank1}) is concave, which is the difference
between a linear function and a convex function. Hence, we need to
construct a convex approximation of constraint (\ref{eq:rank1}).
To address this issue, we introduce the following lemma.

\begin{lemma}\label{lemma-rank-1} Denote by $\mathbf{v}_{max}$
the eigenvector corresponding to the maximum eigenvalue of a matrix
$\mathbf{V}$, we have 
\begin{align*}
\mathrm{Tr}\left\{ \mathbf{v}_{max}\mathbf{v}_{max}^{\mathrm{H}}(\mathbf{Z}-\mathbf{V})\right\}  & =\mathbf{v}_{max}^{\mathrm{H}}\mathbf{Z}\mathbf{v}_{max}-\mathbf{v}_{max}^{\mathrm{H}}\mathbf{V}\mathbf{v}_{max}\\
 & =\mathbf{v}_{max}^{\mathrm{H}}\mathbf{Z}\mathbf{v}_{max}-\lambda_{\max}(\mathbf{V})\\
 & \leq\lambda_{\max}(\mathbf{Z})-\lambda_{\max}(\mathbf{V})
\end{align*}
for any Hermitian matrix $\mathbf{Z}$. \hspace{12cm}$\blacksquare$
\end{lemma}Let $\mathbf{d}_{\mathbf{W}}^{(n)}$ be the eigenvector
corresponding to the maximum eigenvalue of the feasible point $\mathbf{W}^{(n)}$.
Then, by using Lemma \ref{lemma-rank-1}, the surrogate convex constraint
of (\ref{eq:rank1}) is given by 
\begin{align}
\mathrm{Tr}\left\{ \mathbf{W}\right\}   -\lambda_{\max}(\mathbf{W}^{(n)}) -\mathrm{Tr}\left\{ \mathbf{d}_{\mathbf{W}}^{(n)}\mathbf{d}_{\mathbf{W}}^{(n),\mathrm{H}}(\mathbf{W}-\mathbf{W}^{(n)})\right\} \leq\varepsilon.\label{Pro:W-c-1}
\end{align}

Now, we consider constraint (\ref{eq:rank2}). By appying the first-order
Tayler approximation to $||\mathbf{w}||^{2}$, we obtain the following
convex approximation of the constraint in (\ref{eq:rank2}) as\begin{subequations}\label{Pro:W-c-2}
\begin{align}
||\mathbf{w}||^{2}-\mathrm{Tr}\left\{ \mathbf{W}\right\}  & \leq\varepsilon,\label{eq:rank3-1}\\
2\mathrm{Re}\left\{ \mathbf{w}^{(n),\mathrm{H}}\mathbf{w}\right\} -||\mathbf{w}^{(n)}||^{2}-\mathrm{Tr}\left\{ \mathbf{W}\right\}  & \geq-\varepsilon.
\end{align}
\end{subequations}

Finally, the subproblem w.r.t., $\{\mathbf{z},\mathbf{w},\mathbf{W},\mathbf{t}\}$
is formulated as\begin{subequations}\label{Pro:act-pro-1} 
\begin{align}
\max\limits _{R_{\mathrm{sec}},\mathbf{z},\mathbf{w},\mathbf{W},\mathbf{t},\mathbf{x},\mathbf{y}} & R_{\mathrm{sec}}\\
\textrm{s.t.} & (\ref{eq:BTI-1}),(\ref{eq:BTI-2}),(\ref{eq:p-c}),(\ref{eq:sp-c}),\\
 & (\ref{eq:ph2}),(\ref{Pro:W-c-1}),(\ref{Pro:W-c-2})\\
 & \mathbf{W}\succeq\mathbf{0}.
\end{align}
\end{subequations}Problem (\ref{Pro:act-pro-1}) is an SDP and can
be solved by the CVX tool \cite{CVX2018}.

For given $\{\mathbf{w},a_{K},a_{E},v\}$, Problem (\ref{Pro:act-pro})
with optimization variable $\mathbf{e}$ can be solved by applying
the penalty CCP \cite{GuiTSProbust,Gui-letter,PCCP-boyd} to relax
the unit-modulus constraint (\ref{eq:e-c}). Comparing with the SDR
technique, the penalty CCP method is capable of finding a feasible
solution to meet constraint (\ref{eq:e-c}). In particular, the constraint
of (\ref{eq:e-c}) can be relaxed by\begin{subequations}\label{Pro:e-c-1}
\begin{align}
 & |e_{m}^{[j]}|^{2}-2\mathrm{Re}(e_{m}^{\mathrm{H}}e_{m}^{[j]})\leq b_{m}-1,1\leq m\leq M,\label{eq:rank3-1-1}\\
 & |e_{m}|^{2}\leq1+b_{M+m},1\leq m\leq M,
\end{align}
\end{subequations}where $e_{m}^{[j]}$ is any feasible solution and
$\mathbf{b}=[b_{1},...,b_{2M}]^{\mathrm{T}}$ is slack vector variable.
The proof of (\ref{Pro:e-c-1}) can be found in \cite{GuiTSProbust,Gui-letter}.
Following the penalty CCP framework, the subproblem for optimizing
$\mathbf{e}$ is formulated as\begin{subequations}\label{Pro:act-pro-2}
\begin{align}
\max\limits _{R_{\mathrm{sec}},\mathbf{e},\mathbf{x},\mathbf{y}} & R_{\mathrm{sec}}-\lambda^{[j]}||\mathbf{b}||_{1}\\
\textrm{s.t.} & (\ref{eq:BTI-1}),(\ref{eq:BTI-2}),(\ref{Pro:e-c-1}).
\end{align}
\end{subequations}

Problem (\ref{Pro:act-pro-2}) is an SDP and can be solved by the
CVX tool. The algorithm for finding a feasible solution of $\mathbf{e}$
is summarized in Algorithm \ref{Algorithm-analog}.

\begin{algorithm}
\caption{Penalty CCP optimization for reflection beamforming optimization}
\label{Algorithm-analog} \begin{algorithmic}[1] \REQUIRE Initialize
$\mathbf{e}^{[0]}$, $\gamma^{[0]}>1$, and set $j=0$.

\REPEAT

\IF {$j<J_{max}$}

\STATE Update $\mathbf{e}^{[j+1]}$ by solving Problem (\ref{Pro:act-pro-2});

\STATE Update $\lambda^{[j+1]}=\min\{\gamma\lambda^{[j]},\lambda_{max}\}$;

\STATE $j=j+1$;

\ELSE

\STATE Initialize with a new random $\mathbf{e}^{[0]}$, set up $\gamma^{[0]}>1$
again, and set $j=0$.

\ENDIF

\UNTIL $||\mathbf{b}||_{1}\leq\chi$ and $||\mathbf{e}^{[j]}-\mathbf{e}^{[j-1]}||_{1}\leq\nu$.

\STATE Output $\mathbf{e}^{(n+1)}=\mathbf{e}^{[j]}$.

\end{algorithmic} 
\end{algorithm}

In addition, Problem (\ref{Pro:act-pro}) is convex in $\{R_{\mathrm{sec}},v,\mathbf{x},\mathbf{y}\}$
for given $\{\mathbf{z},\mathbf{w},\mathbf{e},\mathbf{t},a_{K},a_{E},\}$,
and convex in $\{R_{\mathrm{sec}},a_{K},a_{E},\mathbf{x},\mathbf{y}\}$
for given $\{\mathbf{z},\mathbf{w},\mathbf{e},\mathbf{t},v,\}$. Finally,
Problem (\ref{Pro:act-pro}) is addressed under the alternate optimization
(AO) framework containing four subproblems. The convergence of the
AO framework can be guaranteed due to the fact that each subproblem
can obtain a non-decreasing sequence of objective function values.

\section{ED Model II: Passive Eavesdropper Model}

In this section, we focus on the transmission design for the passive
attack, which is more practical and more challenging to address, since
the passive ED can hide itself and its CSI is not known \cite{PLS-1,PLS-2}.
The authors of \cite{AAUC} proposed to exploit the angular information
of the ED to combat its passive attack, which is also applicable here.
In this section, the cascaded LIL channel $\mathbf{H}_{K}$ and the
channel $\mathbf{H}_{\mathrm{IRS}}$ are assumed to be perfect, which
is reasonable due to the fact that the pilot information for channel
estimation for LUs is known at the BS.

\subsection{Average eavesdropping rate Maximization}

The signal received by the ED is formulated as 
\begin{align*}
y_{E} & =\mathbf{h}_{E}^{\mathrm{H}}\mathrm{diag}(\mathbf{e}^{*})\mathbf{H}_{\mathrm{IRS}}\mathbf{w}s+\sigma_{E}^{2}.
\end{align*}

Since the ED is passive, we can only detect the activity of the ED
on the line segment between the BS and LU $K$ without knowing its
exact location. This detection of a passive attack is based on spectrum
sensing \cite{PLS-3}. Hence, the average eavesdropping rate is considered
which can be computed as follows \cite{PLS-4,PLS-5,AAUC}: 
\begin{align}
R_{E}^{av}\left(\mathbf{w},\mathbf{e}\right)=  \frac{1}{D_{K}}\int_{0}^{D_{K}}\mathbb{E}_{\{\mathbf{h}_{E}\}}\Biggl[\frac{1}{2}\log_{2}\biggl(1+ \frac{1}{\sigma_{E}^{2}}\left|\mathbf{h}_{E}^{\mathrm{H}}\mathrm{diag}(\mathbf{e}^{*})\mathbf{H}_{\mathrm{IRS}}\mathbf{w}\right|^{2}\biggr)\Biggr]\mathrm{d}_{D_{E}}.\label{eq:av-r}
\end{align}
With (\ref{eq:av-r}), we formulate the following optimization problem:\begin{subequations}\label{pro:average-1}
\begin{align}
\max\limits _{\mathbf{z},\mathbf{w},\mathbf{e},\mathbf{t}} & \left\{ \min_{\forall k\in\mathcal{K}}R_{k}-R_{E}^{av}\left(\mathbf{w},\mathbf{e}\right)\right\} \\
\textrm{s.t.} & (\ref{eq:p-c})-(\ref{eq:ph-c}).
\end{align}
\end{subequations}

The main challenge to solve Problem (\ref{pro:average-1}) is from
the average eavesdropping rate containing the integration over $D_{E}$
and the expectation over $\mathbf{h}_{E}$. To address this issue,
we use Jensen's inequality to construct an upper bound of $R_{E}^{av}\left(\mathbf{w},\mathbf{e}\right)$
given by 
\begin{align}
  R_{E}^{up}\left(\mathbf{w},\mathbf{e}\right)
 & =\frac{1}{2}\log_{2}\left(1+\frac{\int_{0}^{D_{K}}\mathbb{E}_{\{\mathbf{h}_{E}\}}\left[\left|\mathbf{h}_{E}^{\mathrm{H}}\mathrm{diag}(\mathbf{e}^{*})\mathbf{H}_{\mathrm{IRS}}\mathbf{w}\right|^{2}\right]\mathrm{d}_{D_{E}}}{\sigma_{E}^{2}D_{K}}\right)\nonumber \\
 & =\frac{1}{2}\log_{2}\left(1+\frac{1}{\sigma_{E}^{2}}\mathbf{w}^{\mathrm{H}}\mathbf{H}_{\mathrm{IRS}}^{\mathrm{H}}\mathrm{diag}(\mathbf{e})\mathbf{R}_{E}\mathrm{diag}(\mathbf{e}^{*})\mathbf{H}_{\mathrm{IRS}}\mathbf{w}\right),\label{eq:f}
\end{align}
where $\mathbf{R}_{E}=\frac{1}{D_{K}}\int_{0}^{D_{K}}\mathbb{E}_{\{\mathbf{h}_{E}\}}[\mathbf{h}_{E}\mathbf{h}_{E}^{\mathrm{H}}]\mathrm{d}_{D_{E}}$
which can be computed via one-dimension integration.

According to (\ref{eq:los}), we define\begin{subequations}\label{pro:average-1-1}
\begin{align}
\overline{\mathbf{h}}_{E} & =\sqrt{\varrho_{0}\left(\frac{d_{\mathrm{IRS},E}}{d_{0}}\right)^{-\alpha_{\mathrm{IRS}}}\frac{K_{\mathrm{IRS}}}{1+K_{\mathrm{IRS}}}}\mathbf{h}_{E}^{\mathrm{LOS}},\\
\mathbf{R}_{E} & =\varrho_{0}\left(\frac{d_{\mathrm{IRS},E}}{d_{0}}\right)^{-\alpha_{\mathrm{IRS}}}\frac{1}{1+K_{\mathrm{IRS}}}\mathbf{I}_{M},
\end{align}
\end{subequations}where $\overline{\mathbf{h}}_{E}$ describes the
LoS component and is the mean of channel $\mathbf{h}_{E}$. Moreover,
$\mathbf{R}_{E}$ is a positive semi-definite covariance matrix representing
the spatial correlation characteristics of the NLoS component. Therefore,
we have $\mathbf{h}_{E}\sim\mathcal{CN}(\overline{\mathbf{h}}_{E},\mathbf{R}_{E})$
\cite{emil-los}, and further get 
\begin{align*}
  \mathbb{E}_{\{\mathbf{h}_{E}\}}[\mathbf{h}_{E}\mathbf{h}_{E}^{\mathrm{H}}]&=\left[\mathbf{R}_{E}+\overline{\mathbf{h}}_{E}\overline{\mathbf{h}}_{E}^{\mathrm{H}}\right]\\
 & =\varrho_{0}\left(\frac{d_{\mathrm{IRS},E}}{d_{0}}\right)^{-\alpha_{\mathrm{IRS}}}\cdot\left[\frac{1}{1+K_{\mathrm{IRS}}}\mathbf{I}_{M}+\frac{K_{\mathrm{IRS}}}{1+K_{\mathrm{IRS}}}\mathbf{h}_{E}^{\mathrm{LOS}}(\mathbf{h}_{E}^{\mathrm{LOS}})^{\mathrm{H}}\right].
\end{align*}

\subsection{Proposed Algorithm}

By replacing $R_{E}^{av}\left(\mathbf{w},\mathbf{e}\right)$ with
$R_{E}^{up}\left(\mathbf{w},\mathbf{e}\right)$ in the objective function
of Problem (\ref{pro:average-1}), we have\begin{subequations}\label{pro:average-2}
\begin{align}
\max\limits _{\mathbf{z},\mathbf{w},\mathbf{e},\mathbf{t}} & \left\{ \min_{\forall k\in\mathcal{K}}R_{k}-R_{E}^{up}\right\} \\
\textrm{s.t.} & (\ref{eq:p-c})-(\ref{eq:ph-c}).
\end{align}
\end{subequations}

Problem (\ref{pro:average-2}) is still difficult to solve due to
the non-convex constraints and objective function, as well as the
coupled variables $\mathbf{w}$ and $\mathbf{e}$. Hence, we propose
an MM-based AO method to update $\mathbf{w}$ and $\mathbf{e}$ iteratively.
More specifically, by first fixing $\mathbf{e}$, the non-concave
objective function w.r.t., $\{{\bf z},\mathbf{w},\mathbf{t}\}$ is
replaced by its customized concave surrogate function and then solved
by the CVX. $\{{\bf z},\mathbf{w},\mathbf{t}\}$ are then fixed and
the closed-form solution of $\mathbf{e}$ can be found by constructing
an easy-to-solve surrogate objective function w.r.t $\mathbf{e}$.

The surrogate functions of $R_{k}\left(\mathbf{z},t_{k}\right)$ for
$\forall k\in\mathcal{K}_{-K}$ are given by $\widehat{R}_{k}(\mathbf{z},t_{k}|\mathbf{z}^{(n)},t_{k}^{(n)})=\widetilde{R}_{k}(\mathbf{z},t_{k}|\mathbf{z}^{(n)},t_{k}^{(n)})$
given in (\ref{eq:c}), and those of $R_{K}\left(\mathbf{w},\mathbf{e}\right)$
and $R_{E}\left(\mathbf{w},\mathbf{e}\right)$ are given in the following
lemma by using the first-order Taylor approximation.

\begin{lemma}\label{lower-1-1}

Let $\{\mathbf{w}^{(n)},\mathbf{e}^{(n)}\}$ be any feasible solution,
then $R_{K}\left(\mathbf{w},\mathbf{e}\right)$ is lower bounded by
a concave surrogate function $\widehat{R}_{K}\left(\mathbf{w},\mathbf{e}|\mathbf{w}^{(n)},\mathbf{e}^{(n)}\right)$
defined by 
\begin{align}
\widehat{R}_{K}(\mathbf{w},\mathbf{e}|\mathbf{w}^{(n)},\mathbf{e}^{(n)}) & =\frac{1}{2}\log_{2}\Biggl(1-\frac{q_{K}^{(n)}}{\sigma_{K}^{2}}+2\mathrm{Re}\left\{ \frac{q_{K}}{\sigma_{K}^{2}}\right\} \Biggr),\label{eq:r-k}
\end{align}
where $q_{K}^{(n)}=\left|\mathbf{e}^{(n),\mathrm{H}}\mathbf{H}_{K}\mathbf{w}^{(n)}\right|^{2}$
and $q_{K}=\mathbf{e}^{(n),\mathrm{H}}\mathbf{H}_{K}\mathbf{w}^{(n)}\mathbf{w}^{\mathrm{H}}\mathbf{H}_{K}^{\mathrm{H}}\mathbf{e}$.

Meanwhile, $R_{E}\left(\mathbf{w},\mathbf{e}\right)$ is upper bounded
by a convex surrogate function $\widehat{R}_{E}\left(\mathbf{w},\mathbf{e}|\mathbf{w}^{(n)},\mathbf{e}^{(n)}\right)$
given by 
\begin{align}
\widehat{R}_{E}^{up}\left(\mathbf{w},\mathbf{e}|\mathbf{w}^{(n)},\mathbf{e}^{(n)}\right)=  \frac{1}{2}\log_{2}\left(1+\frac{q_{E}^{(n)}}{\sigma_{E}^{2}}\right) +\frac{q_{E}-q_{E}^{(n)}}{2(\sigma_{E}^{2}+q_{E}^{(n)})\ln2},\label{eq:r-e}
\end{align}
where $q_{E}^{(n)}=\mathbf{w}^{(n),\mathrm{H}}\mathbf{H}_{\mathrm{IRS}}^{\mathrm{H}}\mathrm{diag}(\mathbf{e}^{(n)})\mathbf{R}_{E}\mathrm{diag}(\mathbf{e}^{(n),*})\mathbf{H}_{\mathrm{IRS}}\mathbf{w}^{(n)}$
and $q_{E}=\mathbf{w}^{\mathrm{H}}\mathbf{H}_{\mathrm{IRS}}^{\mathrm{H}}\mathrm{diag}(\mathbf{e})\mathbf{R}_{E}\mathrm{diag}(\mathbf{e}^{*})\mathbf{H}_{\mathrm{IRS}}\mathbf{w}$.

\end{lemma}Furthermore, we have the following proposition.

\begin{proposition}\label{proposition-2}

The functions $\{\widehat{R}_{k},\widehat{R}_{K},\widehat{R}_{E}^{up}\}$
preserve the first-order property of functions $\{R_{k},R_{K},R_{E}\}$,
respectively. Let's take $\widehat{R}_{K}$ and $R_{K}$ as an example

\begin{align*}
\nabla_{\mathbf{w}}\widehat{R}_{K}(\mathbf{w},\mathbf{e}^{(n)}|  \mathbf{w}^{(n)},\mathbf{e}^{(n)})|_{\mathbf{w}=\mathbf{w}^{(n)}}
= & \nabla_{\mathbf{w}}R_{K}\left(\mathbf{w},\mathbf{e}^{(n)}\right)|_{\mathbf{w}=\mathbf{w}^{(n)}},\\
\nabla_{\mathbf{e}}\widehat{R}_{K}(\mathbf{w}^{(n)},\mathbf{e}|  \mathbf{w}^{(n)},\mathbf{e}^{(n)})|_{\mathbf{e}=\mathbf{e}^{(n)}}
= & \nabla_{\mathbf{e}}R_{K}\left(\mathbf{w}^{(n)},\mathbf{e}\right)|_{\mathbf{e}=\mathbf{e}^{(n)}}.
\end{align*}

\end{proposition}

\textbf{\textit{Proof: }}See Appendix B in \cite{AAUC}.\hspace{11cm}$\blacksquare$

Giving $\mathbf{e}$ and combining (\ref{eq:c}), (\ref{eq:r-k}),
(\ref{eq:r-e}) and (\ref{eq:ph2}), the subproblem of optimizing
$\{{\bf z},\mathbf{w},\mathbf{t}\}$ is formulated as\begin{subequations}\label{pro:average-w}
\begin{align}
\max\limits _{\mathbf{z},\mathbf{w},\mathbf{t}} & \left\{ \min_{\forall k\in\mathcal{K}}\widehat{R}_{k}-\widehat{R}_{E}^{up}\right\} \\
\textrm{s.t.} & (\ref{eq:p-c}),(\ref{eq:sp-c}),(\ref{eq:ph2}).
\end{align}
\end{subequations}Introducing auxiliary variable $r$, we can transform
Problem (\ref{pro:average-w}) into\begin{subequations}\label{pro:average-w-1}
\begin{align}
\max\limits _{\mathbf{z},\mathbf{w},\mathbf{t},r} & \left\{ r-\widehat{R}_{E}^{up}\right\} \\
\textrm{s.t.} & (\ref{eq:p-c}),(\ref{eq:sp-c}),(\ref{eq:ph2})\\
 & \widehat{R}_{k}\geq r,\forall k\in\mathcal{K},
\end{align}
\end{subequations} which is convex and can be solved by using CVX.

Giving $\{{\bf z},\mathbf{w},\mathbf{t}\}$ and combining (\ref{eq:c}),
(\ref{eq:r-k}) and (\ref{eq:r-e}), the subproblem w.r.t., $\mathbf{e}$
is formulated as 
\begin{align}
\max\limits _{\mathbf{e}} & \left\{ \min_{\forall k\in\mathcal{K}}\widehat{R}_{k}-\widehat{R}_{E}^{up}\right\} ,\textrm{s.t.}(\ref{eq:e-c}).\label{pro:average-e}
\end{align}
Problem (\ref{pro:average-e}) can be addressed by transforming it
into an SOCP under the penalty CCP method mentioned in Section \ref{Pro:act-solution}.
However, the penalty CCP method needs to solve a series of SOCP problems
which incurs a high computational complexity. In the following, we
aim to derive a low-complexity algorithm containing the closed-form
solution of $\mathbf{e}$.

Denote by $\mathcal{R}=\min_{k=1}^{K-1}\left\{ R_{k}\left(\mathbf{z},t_{k}\right)\right\} $
that is a constant, then the subproblem of Problem (\ref{pro:average-2})
corresponding to the optimization of $\mathbf{e}$ is given by 
\begin{align}
\max\limits _{\mathbf{e}} & \left\{ \min\{\mathcal{R},R_{K}\left(\mathbf{e}\right)\}-R_{E}^{up}\left(\mathbf{e}\right)\right\} ,\textrm{s.t.}(\ref{eq:e-c}).\label{pro:average-e2}
\end{align}

Before solving Problem (\ref{pro:average-e2}), we first consider
the following two subproblems: 
\begin{align}
\mathcal{P}1:\thinspace\thinspace\thinspace\thinspace & \min\limits _{\mathbf{e}}R_{E}^{up}\left(\mathbf{e}\right),\textrm{s.t.}(\ref{eq:e-c}),R_{K}\left(\mathbf{e}\right)\geq\mathcal{R}.\\
\mathcal{P}2:\thinspace\thinspace\thinspace\thinspace & \max\limits _{\mathbf{e}}R_{K}\left(\mathbf{e}\right)-R_{E}^{up}\left(\mathbf{e}\right),\textrm{s.t.}(\ref{eq:e-c}),R_{K}\left(\mathbf{e}\right)\leq\mathcal{R}.
\end{align}

Denote the solutions to $\mathcal{P}1$ and $\mathcal{P}2$ as $\mathbf{e}_{1}^{\#}$
and $\mathbf{e}_{2}^{\#}$, respectively. In addition, let us denote
the objective function value of Problem (\ref{pro:average-e2}) as
$obj(\mathbf{e})$, which is a function of $\mathbf{e}$. If $obj(\mathbf{e}_{1}^{\#})\geq obj(\mathbf{e}_{2}^{\#})$,
then the optimal solution of Problem (\ref{pro:average-e2}) is given
by $\mathbf{e}_{1}^{\#}$. Otherwise, the optimal solution is $\mathbf{e}_{2}^{\#}$.

The solutions to Problem $\mathcal{P}1$ and Problem $\mathcal{P}2$
are given in the following lemma.

\begin{lemma}\label{solution} The optimal solution of $\mathcal{P}1$
is given by 
\begin{equation}
\mathbf{e}_{1}^{\#}=\exp\{\mathrm{j}\arg((\lambda_{\max}(\mathbf{A}_{E})\mathbf{I}-\mathbf{A}_{E}+\varrho_{1}^{opt}\mathbf{A}_{K})\mathbf{e}^{(n)})\},
\end{equation}
where $\mathbf{A}_{E}=\text{(\ensuremath{\mathbf{H}_{\mathrm{IRS}}}\ensuremath{\mathbf{w}\mathbf{w}^{\mathrm{H}}\mathbf{H}_{\mathrm{IRS}}^{\mathrm{H}}})}\odot(\mathbf{R}_{E}^{\mathrm{T}}/\sigma_{E}^{2})$,
$\mathbf{A}_{K}=\mathbf{H}_{K}\mathbf{w}\mathbf{w}^{\mathrm{H}}\mathbf{H}_{K}^{\mathrm{H}}/\sigma_{K}^{2}$
and $\varrho_{1}^{opt}$ is the price introduced by the price mechanism
\cite{Pan2019intelleget}.

The optimal solution of $\mathcal{P}2$ is given by 
\begin{equation}
\mathbf{e}_{2}^{\#}=\exp\{\mathrm{j}\arg(\mathbf{c}+\varrho_{2}^{opt}(\lambda_{\max}(\mathbf{A}_{K})\mathbf{I}-\mathbf{A}_{K})\mathbf{e}^{(n)})\},
\end{equation}
where $\varrho_{2}^{opt}$ is the price and \begin{subequations}\label{para}
\begin{align*}
 & \mathbf{c}=\frac{1+d_{K}^{(n)}}{(1+d_{E}^{(n)})^{2}}(\lambda_{\max}(\mathbf{A}_{E})\mathbf{I}-\mathbf{A}_{E})\mathbf{e}^{(n)}+\frac{\mathbf{A}_{K}\mathbf{e}^{(n)}}{1+d_{E}^{(n)}},\\
 & d_{K}^{(n)}=\mathbf{e}^{(n),\mathrm{H}}\mathbf{A}_{K}\mathbf{e}^{(n)},\thinspace\thinspace\thinspace\thinspace\thinspace\thinspace\thinspace\thinspace\thinspace\thinspace\thinspace\thinspace\thinspace d_{E}^{(n)}=\mathbf{e}^{(n),\mathrm{H}}\mathbf{A}_{E}\mathbf{e}^{(n)}.
\end{align*}
\end{subequations}

\end{lemma}

\textbf{\textit{Proof: }}See Appendix \ref{subsec:The-proof-of-3}.\hspace{12cm}$\blacksquare$

Since $\mathbf{e}_{1}^{\#}$ and $\mathbf{e}_{2}^{\#}$ are the globally
optimal solutions of $\mathcal{P}1$ and $\mathcal{P}2$, respectively,
hence $\mathbf{e}^{\#}$ is the globally optimal solution of Problem
(\ref{pro:average-e2}). The optimal price parameter can be obtained
by using the bisection search method detailed in \cite{Pan2019intelleget}.

According to Proposition \ref{proposition-2} and Theorem 1 in \cite{inner-approximation},
the sequence $\{\mathbf{z}^{(n)},\mathbf{w}^{(n)},\mathbf{t}^{(n)},\mathbf{e}^{(n)}\}_{n=1,2,3,\ldots}$
obtained in each iteration is guaranteed to converge to the set of
stationary points of Problem (\ref{pro:average-2}).

\section{Numerical results and discussions}

In this section, we provide numerical results to evaluate the performance
of the proposed schemes. The results are obtained by using a computer
with a 1.99 GHz i7-8550U CPU and 16 GB RAM. The simulated system setup
is measured with polar coordinates: The BS is located at (0 m, 0 m)
and the IRS is placed at (50 m, 0 m) with elevation angle $\phi=\frac{2\pi}{3}$;
$K$ LUs are randomly and uniformly distributed in an area with $D_{k}\sim\mathcal{U}(20\thinspace\mathrm{m},\thinspace40\thinspace\mathrm{m})$
and $\theta_{k}\sim\mathcal{U}(-\frac{\pi}{2},\thinspace\frac{\pi}{2})$
for $\forall k\in\mathcal{K}$, where $\mathcal{U}$ is the uniform
distribution. The ED is located at $(D_{E},\theta_{K})$ with $D_{E}\in(0,D_{K})$.
The pathloss at the distance of 1 m is $-30$ dB, the pathloss exponents
are set to $\alpha_{\mathrm{BS}}=\alpha_{\mathrm{IRS}}=2.2$ and the
Rician factor is 5. The transmit power budget at the BS is $P_{\max}=30$
dBm and the noise powers are $\{\sigma_{i}^{2}=-100\thinspace\thinspace\mathrm{dBm}\}_{\forall i\in\mathcal{K}_{+E}}$.
For the statistical CSI error model, the variances of \{$\boldsymbol{\bigtriangleup}_{i}^{K},\boldsymbol{\bigtriangleup}_{i}^{E}\}_{\forall i\in\mathcal{K}_{-K}}$
are defined as $\{\varepsilon_{K,i}^{2}=\delta_{K}^{2}||\mathbf{h}_{K}^{*}\odot\mathbf{h}_{i}||_{2}^{2},\varepsilon_{E,i}^{2}=\delta_{E}^{2}||\mathbf{h}_{E}^{*}\odot\mathbf{h}_{i}||_{2}^{2}\}_{\forall i\in\mathcal{K}_{-K}}$,
where $\delta_{K}\in[0,1)$ and $\delta_{E}\in[0,1)$ measure the
relative amount of CSI uncertainties. In addition, the outage probability
of secrecy rate is $\rho=0.05$.

\subsection{Robust secrecy rate in ED Model I}

In order to verify the performance of the proposed outage constrained
beamforming in the AAUC, the case of $N=8$ and $K=5$ is simulated.
For comparison, we also consider the ``Non-robust'' as the benchmark
scheme, in which the estimated cascaded LIL and LIE channels are naively
regarded as perfect channels, resulting in the following problem\begin{subequations}\label{pro:non-robust}
\begin{align}
\max\limits _{\mathbf{z},\mathbf{w},\mathbf{e},\mathbf{t}} & \left\{ \min_{\forall k\in\mathcal{K}}R_{k}-R_{E}\right\} \\
\textrm{s.t.} & (\ref{eq:p-c})-(\ref{eq:ph-c}),
\end{align}
\end{subequations}where $R_{K}\left(\mathbf{e}\right)=\frac{1}{2}\log_{2}(1+|\mathbf{e}^{\mathrm{H}}\widehat{\mathbf{H}}_{K}\mathbf{w}|^{2}/\sigma_{K}^{2})$
and $R_{E}\left(\mathbf{e}\right)=\frac{1}{2}\log_{2}(1+|\mathbf{e}^{\mathrm{H}}\widehat{\mathbf{H}}_{E}\mathbf{w}|^{2}/\sigma_{E}^{2})$.
Problem (\ref{pro:non-robust}) can be solved by using the proposed
low complexity algorithm used to solve Problem (\ref{pro:average-1}).

\begin{figure}
\centering \subfigure[Feasibility rate]{\includegraphics[width=3.5in,height=2.5in]{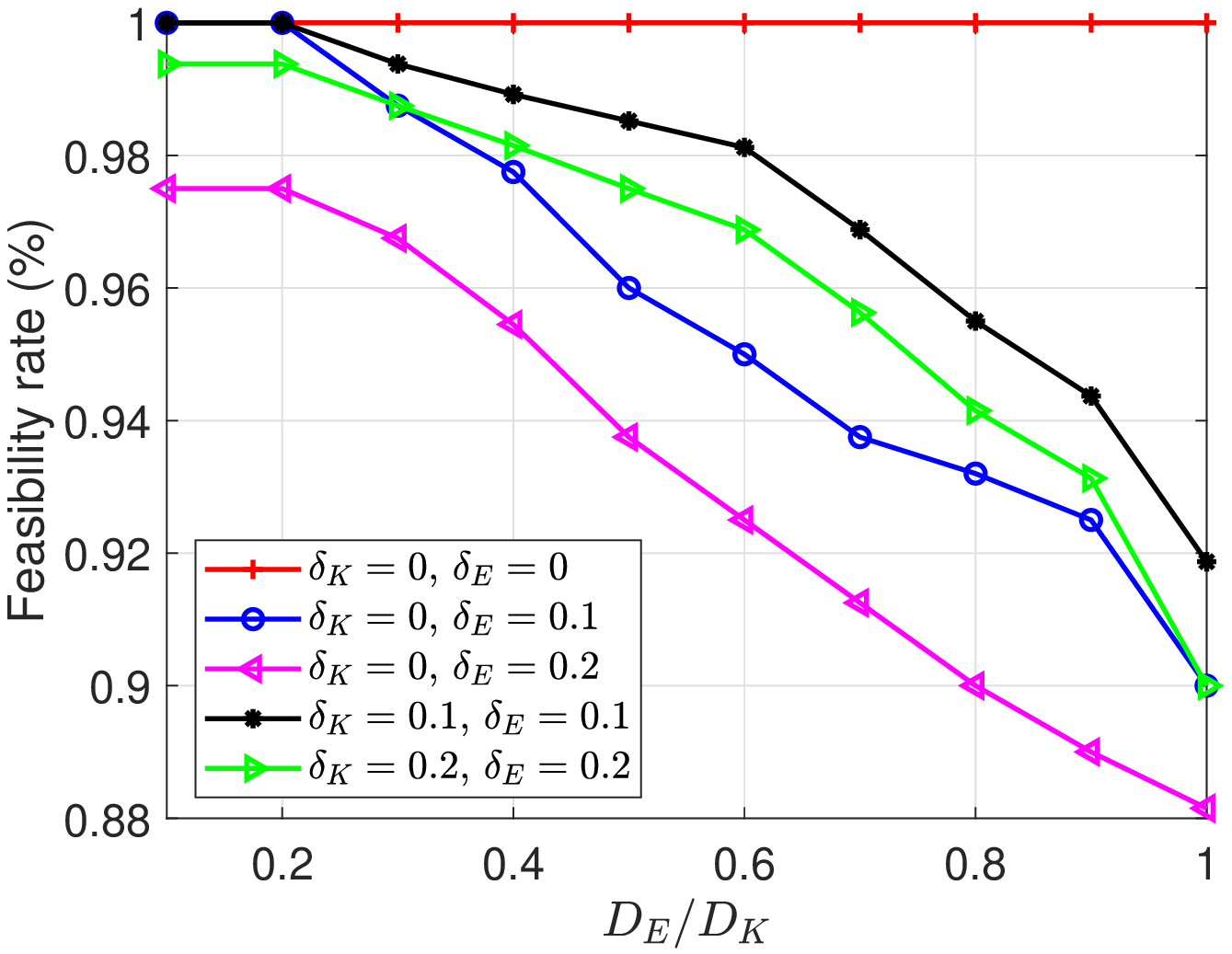}}

\centering \subfigure[Secrecy rate]{\includegraphics[width=3.5in,height=2.5in]{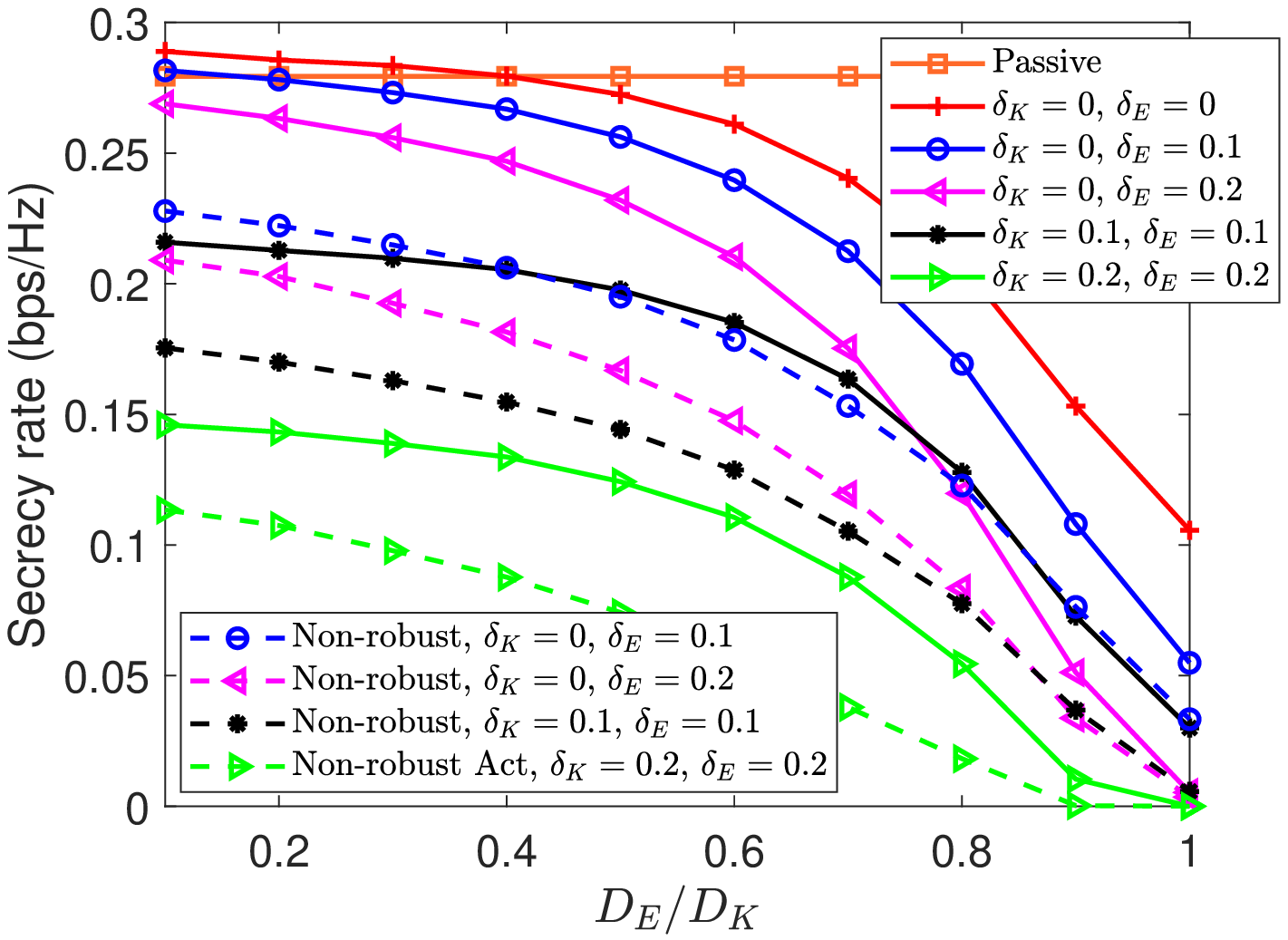}}

\caption{Performance versus $D_{E}/D_{K}$ under $N=8$, $M=32$ and $K=5$.}
\label{act-d} 
\end{figure}

Fig. \ref{act-d} investigates the feasibility rate and the maximum
secrecy rate versus the distance of the ED, in which the coordinate
of X-axis is set to the ratio of $D_{E}/D_{K}$. The feasibility rate
is defined as the ratio of the number of channel realizations that
have a feasible solution to the outage constrained problem of (\ref{pro:outage-1})
to the total number of channel realizations. It is observed from Fig.
\ref{act-d}(a) that the closer the ED is to LU $K$, the lower the
feasibility rate will be, which means that the location of the eavesdropper
imposes a great threat to the security system. From Fig. \ref{act-d}(b),
we can see that the secrecy rate drops fast when the ED approaches
LU $K$, and this secrecy rate reduces to zero when the channel error
is large. At this situation, the whole system is no longer suitable
for secure communication.

\begin{figure}
\centering \subfigure[Feasibility rate]{\includegraphics[width=3.5in,height=2.5in]{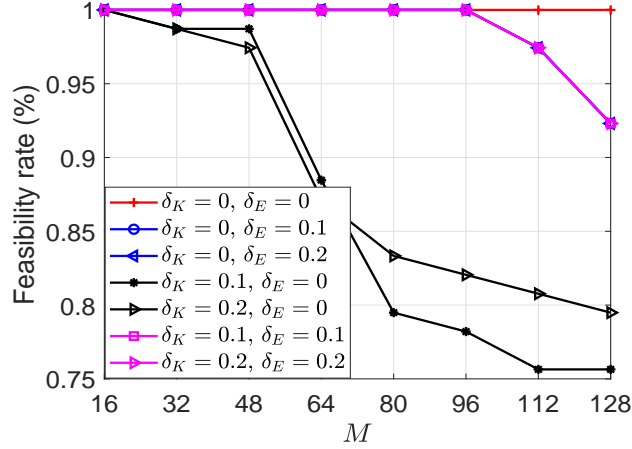}}

\centering \subfigure[Secrecy rate]{\includegraphics[width=3.5in,height=2.5in]{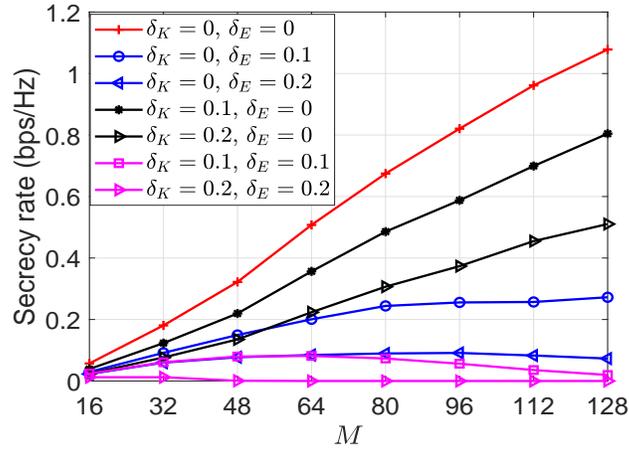}}

\caption{Performance versus $M$ under $N=8$ and $K=5$.}
\label{act-m} 
\end{figure}

Next, the performance versus the size of the IRS, i.e., $M$, is verified
in Fig. \ref{act-m}. Assume that the ED is located at $D_{E}/D_{K}=0.5$.
In Fig. \ref{act-m}(a), the feasibility rate decreases rapidly with
$M$ when the level of channel uncertainty is high. In Fig. \ref{act-m}(b),
the case of $\delta_{K}=\delta_{E}=0$ is regarded as the perfect
cascaded CSI case, and its maximum secrecy rate increases with $M$,
which is consistent with that of Fig. 6 in \cite{Gui2019IRS}. The
performance of $\delta_{K}=\delta_{E}=0$ can be used as the performance
upper bound of the proposed outage constrained beamforming method.
The maximum secrecy rate under small channel uncertainty levels, e.g.,
$(\delta_{K}=\{0.1,0.2\},\delta_{E}=0)$ or $(\delta_{K}=0,\delta_{E}=0.1)$,
also increases with $M$. In addition, it is observed that black lines
of $\delta_{E}=0$ are higher than blue lines of $\delta_{K}=0$,
which means that the negative impact of cascaded LIL channel error
on secrecy rate is smaller than that of the cascaded LIE channel error.
Furthermore, when $\delta_{\mathrm{E}}$ increases to $0.1$ or larger,
secrecy rate starts to decrease at large $M$. The reason is that
increasing $M$ can not only enhance the secrecy rate due to its increased
beamforming gain, but also increase the cascaded channel estimation
error.

Therefore, at small channel uncertainty level, especially that of
the ED, the benefits brought by the increased $M$ outweighs its drawbacks,
and vice versa. This observation provides useful guidance to carefully
chose the size of the IRS according to the level of the channel uncertainty.

\subsection{Average secrecy rate in ED Model II}

This subsection evaluates the performance of the proposed scheme under
the passive attack. In order to evaluate the performance of the proposed
low complexity algorithm, we consider two benchmark algorithms which
are given by: 1) The ``SOCP'' based scheme, i.e., the CVX is used
to solve the SOCP version of Problem (\ref{pro:average-e}). 2) The
``Random'' phase scheme, in which the phases of the reflection elements
are randomly generated.

\begin{figure}
\centering \subfigure[Average secrecy rate]{\includegraphics[width=3.5in,height=2.5in]{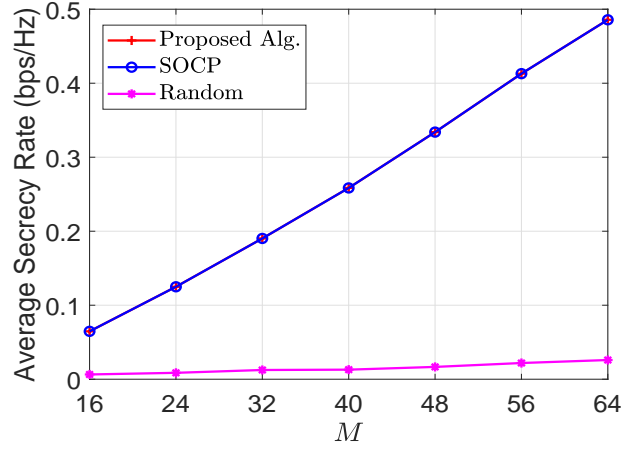}}

\centering \subfigure[Average CPU time]{\includegraphics[width=3.5in,height=2.5in]{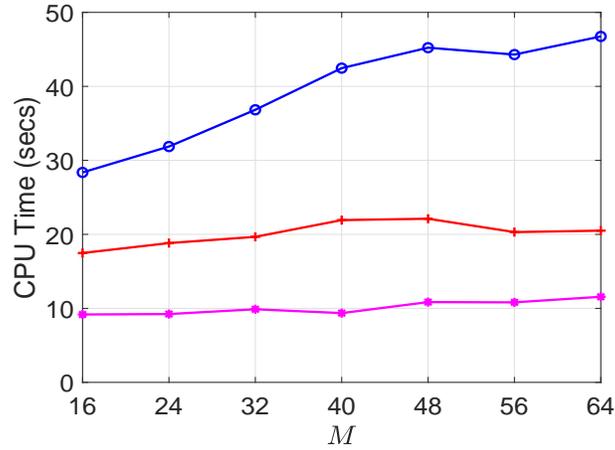}}

\caption{Performance versus $M$ under $N=8$ and $K=5$.}
\label{pass-m} 
\end{figure}

Under the case of $N=8$ and $K=5$, the average secrecy rate versus
$M$ is shown in Fig. \ref{pass-m}(a). It can be seen that the performance
of the proposed algorithm is almost the same as that of the SOCP algorithm,
and both of them outperform the scheme with random phases. Moreover,
increasing the size of the IRS can significantly increase the average
secrecy rate of the system. Accordingly, Fig. \ref{pass-m}(b) describes
the CPU time required for these three algorithms. The simulation setup
is the same as that of Fig. \ref{pass-m}(a). It is observed from
Fig. \ref{pass-m}(b) that the proposed algorthm with closed-form
solution requires much less CPU running time than the SOCP scheme.
In addition, the CPU running time of the SCOP-based algorithm increases
significantly with the size of the IRS, but the proposed algorithm
is not sensitive to the size of the IRS. This is because the computational
complexity of the SOCP depends on $M$, while that of the closed-form
solution does not.

\begin{figure}
\centering \subfigure[Average secrecy rate]{\includegraphics[width=3.5in,height=2.5in]{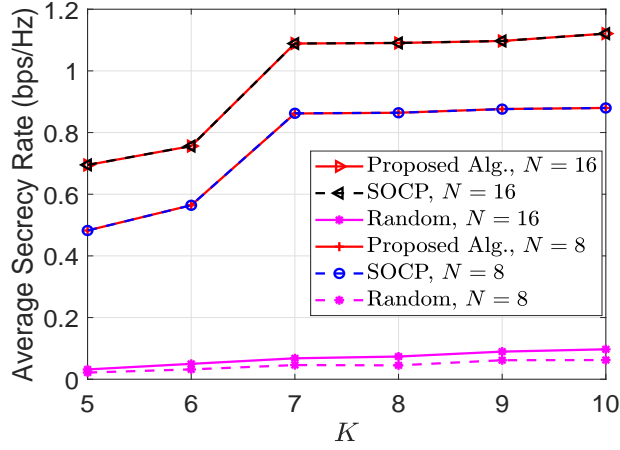}}

\centering \subfigure[Average CPU time]{\includegraphics[width=3.5in,height=2.5in]{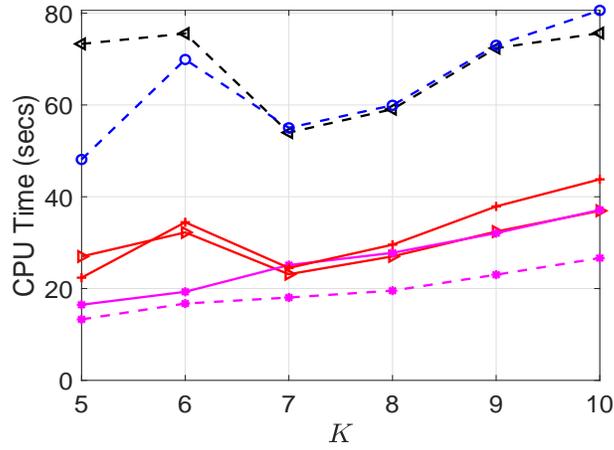}}

\caption{Performance versus $K$ under $M=64$.}
\label{pass-user} 
\end{figure}

Finally, Fig. \ref{pass-user} illustrates the performance and the
complexity versus the number of users in the case of $M=64$. Again,
the proposed algorithm can achieve the same performance of the SOCP-based
scheme with less CPU running time. However, the average secrecy rate
is saturated when the number of users is greater than 7.

\section{Conclusions}

In this work, we have proposed a two-phase IRS-aided communication
system to realize the secure communication under the active attacks
and passive eavesdropping. In order to address the cascaded channel
error caused by the active attacks, we maximize the secrecy rate subject
to secrecy rate outage probability constraints, which has been tackled
by using the Bernstein-Type inequality. For the case of the partial
CSI of the ED, average secrecy rate maximization problem was considered,
which is addressed by the proposed low-complexity algorithm. It was
shown that the negative effect of the ED's channel error is greater
than that of the LU. In addition, the number of elements on the IRS
has a negative impact on system performance when the channel error
is large. This conclusion provides an engineering insight for the
careful selection of the number of elements at the IRS.

\appendices{}

\section{The proof of Lemma \ref{upper}\label{subsec:The-proof-of-1}}

To begin, we introduce the following lemma.

\begin{lemma}\label{wmmse} Let $x\geq0$ be a positive real number,
and consider the function $g_{1}(a,x)=-ax+\ln a+1$, then 
\[
\ln x^{-1}=\max_{a\geq0}g_{1}(a,x).
\]
\end{lemma}By appying Lemma \ref{wmmse}, we can construct an upper
bound of rate $R_{E}\left(\mathbf{w},\mathbf{e}\right)$ as 
\begin{align}
R_{E}\left(\mathbf{w},\mathbf{e}\right) & =\frac{-\ln(1+\left|\mathbf{e}^{\mathrm{H}}\mathbf{H}_{E}\mathbf{w}\right|^{2}/\sigma_{E}^{2})^{-1}}{2\ln2}\nonumber \\
 & \overset{\mathrm{(a)}}{=}\frac{-\max_{a_{E}\geq0}g_{1}(a_{E},1+\left|\mathbf{e}^{\mathrm{H}}\mathbf{H}_{E}\mathbf{w}\right|^{2}/\sigma_{E}^{2})}{2\ln2}\nonumber \\
 & =\frac{\min_{a_{E}\geq0}-g_{1}(a_{E},1+\left|\mathbf{e}^{\mathrm{H}}\mathbf{H}_{E}\mathbf{w}\right|^{2}/\sigma_{E}^{2})}{2\ln2}\nonumber \\
 & \leq\frac{-g_{1}(a_{E},1+\left|\mathbf{e}^{\mathrm{H}}\mathbf{H}_{E}\mathbf{w}\right|^{2}/\sigma_{E}^{2})}{2\ln2}, \textrm{for any }a_{E}\geq0\nonumber \\
 & =\frac{a_{E}\left|\mathbf{e}^{\mathrm{H}}\mathbf{H}_{E}\mathbf{w}\right|^{2}/\sigma_{E}^{2}+a_{E}-\ln a_{E}-1}{2\ln2}\nonumber \\
 & =\widetilde{R}_{E}\left(\mathbf{w},\mathbf{e},a_{E}\right),\label{eq:lower-E-1}
\end{align}
where the equality (a) is from Lemma \ref{wmmse}.

Hence, the proof is completed.

\section{The proof of Lemma \ref{lower-2}\label{subsec:The-proof-of-2}}

To prove Lemma 3, we first introduce the following lemma.

\begin{lemma}\label{wmmse-1} Let $v$ be a complex number, and consider
the function $g_{2}(v,x)=(|x|^{2}+\sigma^{2})|v|^{2}-2\text{Re}\left\{ vx\right\} +1$,
then 
\[
\frac{\sigma^{2}}{|x|^{2}+\sigma^{2}}=\min_{v}g_{2}(a,x).
\]

\end{lemma}

By appying Lemma \ref{wmmse-1}, we can construct a lower bound of
rate $R_{K}\left(\mathbf{w},\mathbf{e}\right)$ as 
\begin{align}
R_{K}\left(\mathbf{w},\mathbf{e}\right) & =\frac{\ln\left(1-\frac{\left|\mathbf{e}^{\mathrm{H}}\mathbf{H}_{K}\mathbf{w}\right|^{2}}{\sigma_{K}^{2}+\left|\mathbf{e}^{\mathrm{H}}\mathbf{H}_{K}\mathbf{w}\right|^{2}}\right)^{-1}}{2\ln2}\nonumber \\
 & \overset{\mathrm{(a)}}{=}\frac{\max_{a_{K}\geq0}g(a_{K},1-\frac{\left|\mathbf{e}^{\mathrm{H}}\mathbf{H}_{K}\mathbf{w}\right|^{2}}{\sigma_{K}^{2}+\left|\mathbf{e}^{\mathrm{H}}\mathbf{H}_{K}\mathbf{w}\right|^{2}})}{2\ln2}\nonumber \\
 & \geq\frac{g(a_{K},1-\frac{\left|\mathbf{e}^{\mathrm{H}}\mathbf{H}_{K}\mathbf{w}\right|^{2}}{\sigma_{K}^{2}+\left|\mathbf{e}^{\mathrm{H}}\mathbf{H}_{K}\mathbf{w}\right|^{2}})}{2\ln2},\textrm{for any }a_{K}\geq0\nonumber \\
 & =\frac{-a_{K}\left(\frac{\sigma_{K}^{2}}{\sigma_{K}^{2}+\left|\mathbf{e}^{\mathrm{H}}\mathbf{H}_{K}\mathbf{w}\right|^{2}}\right)+\ln a_{K}+1}{2\ln2}\nonumber \\
 & \overset{\mathrm{(b)}}{=}\frac{-a_{K}\left(\min_{v}g_{2}(v,\mathbf{e}^{\mathrm{H}}\mathbf{H}_{K}\mathbf{w})\right)+\ln a_{K}+1}{2\ln2}\nonumber \\
 & =\frac{a_{K}\left(\max_{v}-g_{2}(v,\mathbf{e}^{\mathrm{H}}\mathbf{H}_{K}\mathbf{w})\right)+\ln a_{K}+1}{2\ln2}\nonumber \\
 & \geq\frac{a_{K}\left(-g_{2}(v,\mathbf{e}^{\mathrm{H}}\mathbf{H}_{K}\mathbf{w})\right)+\ln a_{K}+1}{2\ln2},\textrm{for any }v\geq0\nonumber \\
 & =\frac{1}{2\ln2}(-a_{K}|v|^{2}|\mathbf{e}^{\mathrm{H}}\mathbf{H}_{K}\mathbf{w}|^{2}-\sigma_{K}^{2}a_{K}|v|^{2}+2a_{K}\text{Re}\left\{ v\mathbf{e}^{\mathrm{H}}\mathbf{H}_{K}\mathbf{w}\right\} -a_{K}+\ln a_{K}+1)\nonumber \\
 & =\widetilde{R}_{K}\left(\mathbf{w},\mathbf{e},a_{K},v\right)\label{eq:lower-K}
\end{align}
where Equality (a) is from Lemma \ref{wmmse}, and Equality (b) is
from Lemma \ref{wmmse-1}.

Hence, the proof is completed.

\section{The proof of Lemma \ref{solution}\label{subsec:The-proof-of-3}}

To begin with, we solve $\mathcal{P}1$: $\mathcal{P}1$ is equivalent
to\begin{subequations}\label{pro:average-p11} 
\begin{align}
\min\limits _{\mathbf{e}} & \thinspace\thinspace\mathbf{e}^{\mathrm{H}}\mathbf{A}_{E}\mathbf{e}\label{eq:ffff}\\
\mathrm{s.t.} & (\ref{eq:e-c}),\\
 & \mathbf{e}^{\mathrm{H}}\mathbf{A}_{K}\mathbf{e}\geq e^{2\mathcal{R}}-1.\label{eq:r-c1}
\end{align}
\end{subequations}

\textit{Step 1: Construct a surrogate problem:} Under the MM algorithm
framework \cite{MM}, we have the following lemma.

\begin{lemma}\label{mm} \cite{mm-1,Pan2019multicell} Given $\mathbf{A}\succeq\mathbf{A}_{0}$
and $\mathbf{x}$, then quadratic function $\mathbf{x}^{\mathrm{H}}\mathbf{A}_{0}\mathbf{x}$
is majorized by $\mathbf{x}^{\mathrm{H}}\mathbf{A}\mathbf{x}-2\mathrm{Re}\{\mathbf{x}^{(n),\mathrm{H}}(\mathbf{A}-\mathbf{A}_{0})\mathbf{x}\}+\mathbf{x}^{(n),\mathrm{H}}(\mathbf{A}-\mathbf{A}_{0})\mathbf{x}^{(n)}$
at $\mathbf{x}^{(n)}$.

\end{lemma}By adopting Lemma \ref{mm} and setting $\mathbf{A}=\lambda_{\max}(\mathbf{A}_{E})\mathbf{I}$
for simplicity, the quadratic objective function in (\ref{eq:ffff})
is majorized by 
\begin{align}
2\lambda_{\max}(\mathbf{A}_{E})M & -2\mathrm{Re}\{\mathbf{e}^{(n),\mathrm{H}}(\lambda_{\max}(\mathbf{A}_{E})\mathbf{I}-\mathbf{A}_{E})\mathbf{e}\} -\mathbf{e}^{(n),\mathrm{H}}\mathbf{A}_{E}\mathbf{e}^{(n)}\label{eq:ff}
\end{align}
at feasible point $\mathbf{e}^{(n)}$. To deal with the non-convex
constraint (\ref{eq:r-c1}), we replace $\mathbf{e}^{\mathrm{H}}\mathbf{A}_{K}\mathbf{e}$
with its linear lower bound, resulting in the following equivalent
constraint 
\begin{equation}
(\ref{eq:r-c1})\Rightarrow2\mathrm{Re}\{\mathbf{e}^{(n),\mathrm{H}}\mathbf{A}_{K}\mathbf{e}\}\geq e^{2\mathcal{R}}-1+\mathbf{e}^{(n),\mathrm{H}}\mathbf{A}_{K}\mathbf{e}^{(n)}.\label{eq:r-c2}
\end{equation}

\textit{Step 2: Closed-form solution :}

By omitting the constant, Problem (\ref{pro:average-p11}) then becomes
\begin{subequations}\label{eq:ave-p12}
\begin{align}
\max\limits _{\mathbf{e}} & \ 2\mathrm{Re}\{\mathbf{e}^{(n),\mathrm{H}}(\lambda_{\max}(\mathbf{A}_{E})\mathbf{I}-\mathbf{A}_{E})\mathbf{e}\}, \\
\textrm{s.t.} & \ (\ref{eq:e-c}),(\ref{eq:r-c2}).
\end{align}
\end{subequations}
According to \cite{Pan2019intelleget}, we introducing a price mechanism
for solving Problem (\ref{eq:ave-p12}), i.e., 
\begin{align*}
\max\limits _{\mathbf{e}} & \thinspace\thinspace2\mathrm{Re}\{\mathbf{e}^{(n),\mathrm{H}}(\lambda_{\max}(\mathbf{A}_{E})\mathbf{I}-\mathbf{A}_{E})\mathbf{e}\} +\varrho_{1}2\mathrm{Re}\{\mathbf{e}^{(n),\mathrm{H}}\mathbf{A}_{K}\mathbf{e}\}\\
\mathrm{s.t.} & \textrm{\thinspace\thinspace(\ref{eq:e-c}).}
\end{align*}
where $\varrho_{1}$ is a nonnegative price. Then, the globally optimal
solution is given by 
\[
\mathbf{e}_{1}^{\#}(\varrho_{1}^{opt})=\exp\{\mathrm{j}\arg((\lambda_{\max}(\mathbf{A}_{E})\mathbf{I}-\mathbf{A}_{E}+\varrho_{1}\mathbf{A}_{K})\mathbf{e}^{(n)})\}.
\]
The optimal $\varrho_{1}^{opt}$ is determined by using the bisection
search method, the detailed information about which can be found in
\cite{Pan2019intelleget}.

Then, we solve $\mathcal{P}2$: $\mathcal{P}2$ is equivalent to\begin{subequations}\label{pro:average-p2}
\begin{align}
\max\limits _{\mathbf{e}} & \ \frac{1+\mathbf{e}^{\mathrm{H}}\mathbf{A}_{K}\mathbf{e}}{1+\mathbf{e}^{\mathrm{H}}\mathbf{A}_{E}\mathbf{e}},\label{eq:obje}\\
\mathrm{s.t.} & \ (\ref{eq:e-c}),\\
 & \ \mathbf{e}^{\mathrm{H}}\mathbf{A}_{K}\mathbf{e}\leq e^{2\mathcal{R}}-1.\label{eq:r-c2-1}
\end{align}
\end{subequations}

\textit{Step 1: Construct a surrogate problem: }Under the MM algorithm
framework, we construct a linear lower bound of the objective funcion
in (\ref{eq:obje}) as 
\begin{align*}
\frac{1+\mathbf{e}^{\mathrm{H}}\mathbf{A}_{K}\mathbf{e}}{1+\mathbf{e}^{\mathrm{H}}\mathbf{A}_{E}\mathbf{e}} & \overset{\mathrm{(a)}}{\geq}\frac{2\mathrm{Re}\{1+d_{K}\}}{1+d_{E}^{(n)}}-\frac{1+d_{K}^{(n)}}{(1+d_{E}^{(n)})^{2}}\left(1+\mathbf{e}^{\mathrm{H}}\mathbf{A}_{E}\mathbf{e}\right)\\
 & \overset{\mathrm{(b)}}{\geq}\frac{2\mathrm{Re}\{1+d_{K}\}}{1+d_{E}^{(n)}}-\frac{1+d_{K}^{(n)}}{(1+d_{E}^{(n)})^{2}}-\frac{1+d_{K}^{(n)}}{(1+d_{E}^{(n)})^{2}}(2\lambda_{\max}(\mathbf{A}_{E})M\\
 & \ -2\mathrm{Re}\{\mathbf{e}^{(n),\mathrm{H}}(\lambda_{\max}(\mathbf{A}_{E})\mathbf{I}-\mathbf{A}_{E})\mathbf{e}\}-d_{E}^{(n)})\\
 & =2\mathrm{Re}\{\mathbf{c}^{\mathrm{H}}\mathbf{e}\}+\mathrm{const},
\end{align*}
where $d_{K}=\mathbf{e}^{(n),\mathrm{H}}\mathbf{A}_{K}\mathbf{e}$.
$\{d_{K}^{(n)},d_{E}^{(n)},\mathbf{c}\}$ are defined in Lemma \ref{solution}.
Inequality (a) is due to Lemma \ref{lower-1}, and inequality (b)
is from Lemma \ref{mm}. By using Lemma \ref{mm} again, the convex
constraint (\ref{eq:r-c2-1}) can be replaced by an easy-to-solve
form as 
\begin{align}
 & (\ref{eq:r-c2-1})\Rightarrow2\mathrm{Re}\{\mathbf{e}^{(n),\mathrm{H}}(\lambda_{\max}(\mathbf{A}_{K})\mathbf{I}-\mathbf{A}_{K})\mathbf{e}\}\nonumber \\
 & \geq-2\lambda_{\max}(\mathbf{A}_{K})M-e^{2\mathcal{R}}+1-\mathbf{e}^{(n),\mathrm{H}}\mathbf{A}_{K}\mathbf{e}^{(n)}.\label{eq:ff-1}
\end{align}

\textit{Step 2: Closed-form solution: }By omitting the constant, Problem
(\ref{pro:average-p2}) is then equivalent to 
\begin{subequations}
\begin{align}
\max\limits _{\mathbf{e}} & \thinspace\thinspace2\mathrm{Re}\{\mathbf{c}^{\mathrm{H}}\mathbf{e}\},\\
\mathrm{s.t.}& \ (\ref{eq:e-c}),(\ref{eq:ff-1}).
\end{align}
\end{subequations}
By using the price mechanism, Problem (\ref{pro:average-p2}) is reformulated
as 
\begin{align*}
\max\limits _{\mathbf{e}} & \thinspace\thinspace2\mathrm{Re}\{\mathbf{c}^{\mathrm{H}}\mathbf{e}\}+\varrho_{2}2\mathrm{Re}\{\mathbf{e}^{(n),\mathrm{H}}(\lambda_{\max}(\mathbf{A}_{K})\mathbf{I}-\mathbf{A}_{K})\mathbf{e}\}\\
\mathrm{s.t.} & (\ref{eq:e-c}).
\end{align*}
where $\varrho_{2}$ is a nonnegative price. The globally optimal
solution is given by $\mathbf{e}_{2}^{\#}(\varrho_{2}^{opt})=\exp\{\mathrm{j}\arg(\mathbf{c}+\varrho_{2}(\lambda_{\max}(\mathbf{A}_{K})\mathbf{I}-\mathbf{A}_{K})\mathbf{e}^{(n)})\}$
where the optimal $\varrho_{2}^{opt}$ is determined by using the
bisection search method.

Hence, the proof is completed.

 \bibliographystyle{IEEEtran}
\bibliography{bibfile}

\begin{thebibliography}{10}
\providecommand{\url}[1]{#1}
\csname url@samestyle\endcsname
\providecommand{\newblock}{\relax}
\providecommand{\bibinfo}[2]{#2}
\providecommand{\BIBentrySTDinterwordspacing}{\spaceskip=0pt\relax}
\providecommand{\BIBentryALTinterwordstretchfactor}{4}
\providecommand{\BIBentryALTinterwordspacing}{\spaceskip=\fontdimen2\font plus
\BIBentryALTinterwordstretchfactor\fontdimen3\font minus
  \fontdimen4\font\relax}
\providecommand{\BIBforeignlanguage}[2]{{%
\expandafter\ifx\csname l@#1\endcsname\relax
\typeout{** WARNING: IEEEtran.bst: No hyphenation pattern has been}%
\typeout{** loaded for the language `#1'. Using the pattern for}%
\typeout{** the default language instead.}%
\else
\language=\csname l@#1\endcsname
\fi
#2}}
\providecommand{\BIBdecl}{\relax}
\BIBdecl

\bibitem{cryptography1976}
W.~Diffie and M.~E. Hellman, ``{New directions in cryptography},'' \emph{IEEE
  Trans. Inf. Theory}, vol. IT-22, no.~6, pp. 644--654, Nov. 1976.

\bibitem{Wyner1975}
A.~D. Wyner, ``{The wire-tap channel},'' \emph{Bell Syst. Tech. J.}, vol.~54,
  no.~8, pp. 1355--1387, Oct. 1975.

\bibitem{Marco-4}
E.~{Basar}, M.~{Di~Renzo}, J.~{De Rosny}, M.~{Debbah}, M.~{Alouini}, and
  R.~{Zhang}, ``Wireless communications through reconfigurable intelligent
  surfaces,'' \emph{IEEE Access}, vol.~7, pp. 116\,753--116\,773, 2019.

\bibitem{Marco-3}
\BIBentryALTinterwordspacing
K.~Ntontin, M.~{Di~Renzo}, J.~Song \emph{et~al.}, ``{Reconfigurable intelligent
  surfaces vs. relaying: Differences, similarities, and performance
  comparison},'' 2019. [Online]. Available:
  \url{https://arxiv.org/abs/1908.08747}
\BIBentrySTDinterwordspacing

\bibitem{KK2020}
\BIBentryALTinterwordspacing
K.-K. Wong, K.-F. Tong, Z.~Chu, and Y.~Zhang, ``{A vision to smart radio
  environment: Surface wave communication superhighways},'' 2020. [Online].
  Available: \url{https://arxiv.org/abs/2005.14082}
\BIBentrySTDinterwordspacing

\bibitem{Shen2019secrecy}
H.~{Shen}, W.~{Xu}, S.~{Gong}, Z.~{He}, and C.~{Zhao}, ``{Secrecy rate
  maximization for intelligent reflecting surface assisted multi-antenna
  communications},'' \emph{IEEE Commun. Lett.}, vol.~23, no.~9, pp. 1488--1492,
  Sept. 2019.

\bibitem{sheng2020}
S.~{Hong}, C.~{Pan}, H.~{Ren}, K.~{Wang}, and A.~{Nallanathan},
  ``Artificial-noise-aided secure mimo wireless communications via intelligent
  reflecting surface,'' \emph{IEEE Trans. Commun.}, vol.~68, no.~12, pp.
  7851--7866, Dec. 2020.

\bibitem{Guan-sec}
\BIBentryALTinterwordspacing
X.~Guan, Q.~Wu, and R.~Zhang, ``{Intelligent reflecting surface assisted
  secrecy communication: Is artificial noise helpful or not?}'' 2020. [Online].
  Available: \url{https://arxiv.org/abs/1907.12839}
\BIBentrySTDinterwordspacing

\bibitem{yu-robust}
\BIBentryALTinterwordspacing
X.~Yu, D.~Xu, Y.~Sun, D.~W.~K. Ng, and R.~Schober, ``{Robust and secure
  wireless communications via intelligent reflecting surfaces},'' 2020.
  [Online]. Available: \url{https://arxiv.org/abs/1912.01497}
\BIBentrySTDinterwordspacing

\bibitem{Hong-robust}
S.~Hong, C.~Pan, H.~Ren, K.~Wang, K.~K. Chai, and A.~Nallanathan, ``{Robust
  transmission design for intelligent reflecting surface aided secure
  communication systems with imperfect cascaded CSI},'' \emph{IEEE Trans.
  Wireless Commun., early access}, pp. 1--1, 2020.

\bibitem{PLS-1}
D.~{Kapetanovic}, G.~{Zheng}, and F.~{Rusek}, ``{Physical layer security for
  massive MIMO: An overview on passive eavesdropping and active attacks},''
  \emph{IEEE Commun. Mag.}, vol.~53, no.~6, pp. 21--27, Jun. 2015.

\bibitem{PLS-2}
Y.~{Wu}, A.~{Khisti}, C.~{Xiao}, G.~{Caire}, K.~{Wong}, and X.~{Gao}, ``A
  survey of physical layer security techniques for 5g wireless networks and
  challenges ahead,'' \emph{IEEE J. Sel. Areas Commun.}, vol.~36, no.~4, pp.
  679--695, Apr. 2018.

\bibitem{GuiTSProbust}
G.~{Zhou}, C.~{Pan}, H.~{Ren}, K.~{Wang}, and A.~{Nallanathan}, ``{A framework
  of robust transmission design for IRS-aided MISO communications with
  imperfect cascaded channels},'' \emph{IEEE Trans. Signal Process.}, vol.~68,
  pp. 5092--5106, Aug. 2020.

\bibitem{Gui2019IRS}
------, ``{Intelligent reflecting surface aided multigroup multicast MISO
  communication systems},'' \emph{IEEE Trans. Signal Process.}, pp. 1--1, 2020.

\bibitem{AAUC}
S.~Wang, M.~Wen, M.~Xia, R.~Wang, Q.~Hao, and Y.-C. Wu.

\bibitem{book-convex}
S.~Boyd and L.~Vandenberghe, \emph{{Convex optimization}}.\hskip 1em plus 0.5em
  minus 0.4em\relax Cambridge Univ. Press, 2004.

\bibitem{BTI-II}
\BIBentryALTinterwordspacing
I.~Bechar, ``{A bernstein-type inequality for stochastic processes of quadratic
  forms of Gaussian variables},'' 2009. [Online]. Available:
  \url{https://arxiv.org/abs/0909.3595}
\BIBentrySTDinterwordspacing

\bibitem{CVX2018}
M.~Grant and S.~Boyd, ``{CVX: MATLAB software for disciplined convex
  programming},'' \emph{Version 2.1. [Online] http://cvxr.com/cvx, Dec. 2018.}

\bibitem{Gui-letter}
G.~Zhou, C.~Pan, H.~Ren \emph{et~al.}, ``{Robust beamforming design for
  intelligent reflecting surface aided MISO communication systems},''
  \emph{IEEE Wireless Commun. Lett.}, vol.~9, no.~10, pp. 1658--1662, Oct.
  2020.

\bibitem{PCCP-boyd}
\BIBentryALTinterwordspacing
T.~Lipp and S.~Boyd, ``{Variations and extension of the convex-concave
  procedure},'' \emph{Optim. Eng.}, vol.~17, no.~2, pp. 263--287, 2016.
  [Online]. Available: \url{https://doi.org/10.1007/s11081-015-9294-x}
\BIBentrySTDinterwordspacing

\bibitem{PLS-3}
A.~Chaman, J.~Wang, J.~Sun, H.~Hassanieh, and R.~R. Choudhury, ``{Ghostbuster:
  Detecting the presence of hidden eavesdroppers},'' \emph{Proc. ACM Mobicom},
  pp. 337--351.

\bibitem{PLS-4}
A.~{Mukherjee}, ``{Physical-layer security in the internet of things: Sensing
  and communication confidentiality under resource constraints},'' \emph{Proc.
  IEEE}, vol. 103, no.~10, pp. 1747--1761, 2015.

\bibitem{PLS-5}
L.~{Mucchi}, L.~{Ronga}, X.~{Zhou}, K.~{Huang}, Y.~{Chen}, and R.~{Wang}, ``{A
  new metric for measuring the security of an environment: The secrecy
  pressure},'' \emph{IEEE Trans. Wireless Commun.}, vol.~16, no.~5, pp.
  3416--3430, 2017.

\bibitem{emil-los}
Ö.~{Özdogan}, E.~{Björnson}, and E.~G. {Larsson}, ``{Massive MIMO with
  spatially correlated rician fading channels},'' \emph{IEEE Trans. Commun.},
  vol.~67, no.~5, pp. 3234--3250, May 2019.

\bibitem{Pan2019intelleget}
C.~Pan, H.~Ren, K.~Wang, M.~Elkashlan, A.~Nallanathan, J.~Wang, and L.~Hanzo,
  ``{Intelligent reflecting surface aided MIMO broadcasting for simultaneous
  wireless information and power transfer},'' \emph{IEEE J. Sel. Areas
  Commun.}, 2020.

\bibitem{inner-approximation}
B.~R. Marks and G.~P. Wright, ``A general inner approximation algorithm for
  nonconvex mathematical programs,'' \emph{Operation Research}, vol.~26, no.~4,
  Jul. 1978.

\bibitem{MM}
Y.~Sun, P.~Babu, and D.~P. Palomar, ``Majorization-minimization algorithms in
  signal processing, communications, and machine learning,'' \emph{IEEE Trans.
  Signal Process.}, vol.~65, no.~3, pp. 794--816, Feb. 2017.

\bibitem{mm-1}
J.~{Song}, P.~{Babu}, and D.~P. {Palomar}, ``Optimization methods for designing
  sequences with low autocorrelation sidelobes,'' \emph{IEEE Trans. Signal
  Process.}, vol.~63, no.~15, pp. 3998--4009, Aug. 2015.

\bibitem{Pan2019multicell}
C.~{Pan}, H.~{Ren}, K.~{Wang}, W.~{Xu}, M.~{Elkashlan}, A.~{Nallanathan}, and
  L.~{Hanzo}, ``{Multicell MIMO communications relying on intelligent
  reflecting surfaces},'' \emph{IEEE Trans. Wireless Commun.}, pp. 1--1, 2020.

\end{thebibliography}

\end{document}